\documentclass[aos]{imsart}

\usepackage[utf8]{inputenc}
\usepackage{natbib,hyperref}
\usepackage{amsbsy,amsfonts,amsmath,amssymb,amsthm,bm,bbm,latexsym,mathabx}
\usepackage{graphicx,epsfig,xcolor}
\usepackage{algorithm,algpseudocode}
\usepackage{changes}
\usepackage{comment}

\startlocaldefs

\theoremstyle{plain}

\newtheorem{lemma}{Lemma}
\newtheorem{theorem}{Theorem}
\newtheorem{corollary}{Corollary}

\theoremstyle{definition}

\newtheorem{remark}{Remark}

\def\bse{\begin{eqnarray*}}
\def\ese{\end{eqnarray*}}
\def\be{\begin{eqnarray}}
\def\ee{\end{eqnarray}}
\def\n{\nonumber}

\def\sumi{\sum_{i=1}^n}

\def\sumI{\sum_{i=1}^N}
\def\sumJ{\sum_{j=1}^N}

\def\wh{\widehat}
\def\wt{\widetilde}

\def\dag{\dagger}
\def\trans{^\mathrm{T}}

\def\I{\mathbb{I}}
\def\E{\mathrm{E}}
\def\Ep{\mathrm{E}_p}
\def\Eq{\mathrm{E}_q}
\def\bias{\mathrm{bias}}

\def\var{\mathrm{var}}

\def\eff{_{\rm eff}}

\DeclareMathOperator*{\argmin}{argmin}

\def\ba{{\boldsymbol\alpha}}
\def\bb{{\boldsymbol\beta}}

\def\bt{{\boldsymbol\theta}}

\def\bphi{{\boldsymbol\phi}}

\def\a{{\mathbf a}}
\def\A{{\mathbf A}}
\def\b{{\mathbf b}}

\def\bI{{\mathbf I}}

\def\calI{\mathcal{I}}
\def\calP{\mathcal{P}}
\def\calQ{\mathcal{Q}}

\def\bbR{{\mathbb R}}

\def\U{{\mathbf U}}
\def\v{{\mathbf v}}

\def\x{{\mathbf x}}
\def\X{{\mathbf X}}

\def\0{{\mathbf 0}}

\def\fxy{f_{\X\mid Y}}

\def\py{p_{Y}}
\def\pyx{p_{Y\mid\X}}
\def\pxy{p_{\X\mid Y}}

\def\qy{q_{Y}}

\def\qxy{q_{\X\mid Y}}

\def\boxit#1{\vbox{\hrule\hbox{\vrule\kern6pt\vbox{\kern6pt#1\kern6pt}\kern6pt\vrule}\hrule}}

\endlocaldefs

\begin{document}

\begin{frontmatter}
\title{Efficient Inference under Label Shift in Unsupervised Domain Adaptation}

\begin{aug}
\author[A]{\fnms{Seong-ho}~\snm{Lee}
\ead[label=e1]{seongho@uos.ac.kr}},
\author[B]{\fnms{Yanyuan}~\snm{Ma}
\ead[label=e2]{yzm63@psu.edu}}
\and
\author[C]{\fnms{Jiwei}~\snm{Zhao}
\ead[label=e3]{jiwei.zhao@wisc.edu}}

\address[A]{Department of Statistics, University of Seoul, Seoul, South Korea\printead[presep={ ,\ }]{e1}}
\address[B]{Department of Statistics, Pennsylvania State University, PA, USA\printead[presep={ ,\ }]{e2}}
\address[C]{Departments of Statistics and of Biostatistics \& Medical Informatics, University of Wisconsin-Madison, WI, USA\printead[presep={ ,\ }]{e3}}
\end{aug}

\begin{abstract}
In many real-world applications, researchers aim to deploy models trained in a source domain to a target domain, where obtaining labeled data is often expensive, time-consuming, or even infeasible. 
While most existing literature assumes that the labeled source data and the unlabeled target data follow the same distribution, distribution shifts are common in practice. 
This paper focuses on label shift and develops efficient inference procedures for general parameters characterizing the unlabeled target population. 
A central idea is to model the outcome density ratio between the labeled and unlabeled data.
To this end, we propose a progressive estimation strategy that unfolds
in three stages: an initial heuristic guess, a consistent estimation,
and ultimately, an efficient estimation.  
This self-evolving process is novel in the statistical literature and
of independent interest.  
We also highlight the connection between our approach and prediction-powered inference (PPI), which uses machine learning models to improve statistical inference in related settings.
We rigorously establish the asymptotic properties of the proposed estimators and demonstrate their superior performance compared to existing methods. 
Through simulation studies and multiple real-world applications, we illustrate both the theoretical contributions and practical benefits of our approach.
\end{abstract}

\begin{keyword}
\kwd{distribution shift}
\kwd{efficient estimation}
\kwd{efficient influence function}
\kwd{label shift}
\kwd{semiparametric statistics}
\end{keyword}

\end{frontmatter}



\section{Introduction}\label{sec:intro}

In many real-world applications, researchers are interested in
deploying models and algorithms trained on a source domain to a target
domain where obtaining labeled data is often expensive,
time-consuming, or even infeasible. 
For example, in healthcare, disease prediction models trained on data from one
hospital may be applied to another 
population of patients at risk but not yet diagnosed.
In social science, models are often evaluated for their
generalizability across survey populations that might differ
demographically or culturally, for which the outcome of interest is not fully collected and time-consuming to collect. In autonomous driving, models trained on simulated data or
limited environment must adapt to new, unseen
locations. 
In the literature, this type of research problems was termed
unsupervised domain adaptation (also known as transductive transfer
learning); see an excellent review paper by \cite{kouw2019review}. 

 In such scenarios where acquiring gold-standard labels in the
  target domain is costly and labor-intensive, machine learning
  (ML)-based algorithms and models offer an efficient solution,
  substantially reducing the cost of outcome collection and
  accelerating scientific research across disciplines
  \citep{wang2023scientific}.
For instance, in electronic health records (EHR), true
disease statuses are accessible through expert-led manual chart
review, and this manual review requires substantial costs due to the large scale nature of EHR.
As a cost-effective solution, ML models can provide approximation of the disease statuses with accuracy
comparable to manual review, thereby streamlining downstream clinical
analyses \citep{yang2022large, zhou2022cancerbert}.  
Similarly, although population biobanks provide extensive data for
high-throughput genetic research, many phenotypes and disease outcomes
remain unmeasured due to prohibitive costs. Resolving this practical issue, ML approaches have been
proven effective in facilitating novel genetic discoveries and
advancing scientific inquiry \citep{an2023deep, mccaw2024synthetic,
  miao2024valid}. 

However, the inherent complexity and the use of
ML predictions pose significant challenges in  statistical
inferences. Using such predictions without recognizing their
differences from observed gold-standard data can result in biased
outcomes and misleading scientific conclusions.
To address this, researchers have introduced methods that fuse
extensive ML predictions with limited gold-standard data to ensure the
validity of ML-assisted statistical inference.
Over the past five years, this topic has gained significant importance
and popularity in the literature.
For easier exposition, consider the availability of both labeled data
\be\label{eq:data:P}
(Y_i,\X_i), i=1,\dots,n, \mbox{ drawn as a random sample from source 
population } \calP,
\ee
and unlabeled data
\be\label{eq:data:Q=P}
\X_i, i=n+1,\dots,N, \mbox{ drawn as a random sample from target 
population } \calQ,
\ee
where the assumption $\calP = \calQ$ is tentatively adopted for simplicity.
\cite{wang2020methods} proposed a \emph{post-prediction inference} procedure, and
\cite{motwani2023revisiting} conducted a more rigorous investigation
on this \emph{inference after prediction} procedure, and showed that,
the method proposed in \cite{wang2020methods} could only provide a
valid inference in some restrictive scenarios in the context of least
squares estimation.
More recently, \cite{angelopoulos2023prediction} proposed the approach
termed \emph{prediction-powered inference} (PPI), which yields
valid inference  even when the ML predictions were of a low quality.
To illustrate, PPI estimates the outcome mean $\theta=\E(Y)$ as
\be\label{eq:PPI}
\wh\theta &=& \underbrace{\frac1{N-n}\sum_{i=n+1}^{N} \wh
  y_i}_{\substack{\text{Direct estimation using}\\ \text{predictions
      on unlabeled data}}}
\quad +
\underbrace{\frac1n\sumi (y_i-\wh y_i)}_{\substack{\text{Rectifier:
      measure the bias of using}\\ \text{predictions on labeled
      data}}},
\ee
where $\wh y_i$'s, representing the conditional mean $\E(Y\mid \x_i)$,
are the predicted outcomes produced by the ML algorithm/model.
A few extensions of PPI were also proposed in the
past two years, including but not limited to, \cite{zrnic2024cross}
which considered the availability of the ML predictions, and
\cite{angelopoulos2023ppi++, miao2023assumption, gronsbell2024another}
which investigated the statistical efficiency gains from different
perspectives.

It is worthwhile to note, in the statistical literature, terms as \emph{semi-supervised inference}, researchers have also proposed different methods of utilizing the unlabeled data for different purposes \citep{chakrabortty2018, yuval2022semi, azriel, livne2022improved, hou2023surrogate, song2023general}, including settings under high dimensionality \citep{zhang2019, caiexplained, 2019jelena, deng2023optimal}.

Despite significant progress in conducting inference under such a scenario, several key aspects remain unclear in the literature.
This paper is motivated by the challenge of distribution shift \citep{quinonero2009dataset} between source domain labeled data \eqref{eq:data:P} and target domain unlabeled data \eqref{eq:data:Q=P}, in the sense that the unlabeled data were collected under conditions different from those of the labeled data.
The scenario $\calP\neq \calQ$ is more realistic and more applicable in many practical settings.
Indeed, various forms of distribution shift have been studied extensively in the literature, including covariate shift (the conditional distribution of $Y$ given $\X$ remains the same while the marginal distribution of $\X$ shifts), and label shift (the conditional distribution of $\X$ given $Y$ remains the same while the marginal distribution of $Y$ shifts).

The goal of this paper is to develop an efficient procedure for estimation and inference under \emph{label shift}, where \emph{efficiency} is defined in the context of statistical estimation \citep{bickel1993efficient, tsiatis2006semiparametric}.
It is worth noting that in the original PPI paper, \cite{angelopoulos2023prediction} addressed scenarios involving distribution shift.
However, under label shift, their method is limited to cases with discrete $Y$, similar to the approaches in \cite{lipton2018detecting} and \cite{tian2023elsa}.
Moreover, estimation efficiency was not considered in their work.
To the best of our knowledge, efficient estimation with label shift
under this setting has not been explored in the literature, despite
the availability of various methods proposed in both statistical and
machine learning contexts \citep{storkey2009training, zhang2013domain,
  du2014semi, iyer2014maximum, nguyen2016continuous, tasche2017fisher,
  kim2024retasa, lee2024doubly}. 
We also emphasize that the setting considered in this paper falls under the
broad umbrella of unsupervised domain adaptation,
where the outcome $Y$ \emph{is not} observed in the unlabeled dataset
$\calQ$. 
This is fundamentally different from settings where $Y$ \emph{is} observed in $\calQ$, as in \cite{li2023efficient} and \cite{qiu2024efficient}, for reasons discussed below.

Under label shift, the most challenging step is identifying the
distribution discrepancy, characterized by the density
  ratio $\rho(y)$ defined
in \eqref{eq:define:rho} in Section~\ref{sec:review}.
In the setting of unsupervised domain adaptation, this step is particularly challenging due to the absence of outcome data $Y$ from the population $\calQ$.
This absence is the key distinction from other settings pointed out in the previous paragraph. 
Under this setting, \cite{lee2024doubly} proposed estimation procedures
for the general parameter $\bt$ in the unlabeled data population $\calQ$
defined in \eqref{eq:theta0}, strategically avoiding
the challenge of identifying $\rho(y)$.
These procedures guarantee consistency but not efficiency,
as the efficiency is achieved only if a heuristic guess on $\rho(y)$ is correct,
which is almost impossible due to the absence of $Y$ in $\calQ$.
In contrast, our work offers a  way to guarantee efficiency in
estimating $\bt$ through a novel way of estimating $\rho(y)$.
A critical insight enabling the estimation of $\rho(y)$ is that, 
although $\rho(y)$ is a \emph{local} feature of $\calQ$, 
its estimation can be reframed as a problem of estimating a
\emph{global} feature of $\calQ$. 
Leveraging this insight, we propose estimation procedures for $\rho(y)$ 
which, in turn, enable efficient estimation of $\bt$ in $\calQ$.
For estimating $\rho(y)$ itself, we also introduce a progressive
estimation process, evolving through three stages: an initial
heuristic guess, a consistent estimation, and ultimately an efficient
estimation.
Different from the traditional one-step estimation
\citep{bickel1975one} commonly used in semiparametric statistics
\citep{bickel1993efficient} and targeted learning
\citep{van2011targeted}, 
our estimation is multi-step, where
  in each step, we adopt the same procedure to obtain an improved estimate
based on the current estimate.
This self-driven evolutionary process is not common in the statistical
literature and might be of independent interest.
The detailed methodology underlying our proposal is presented in
Section~\ref{sec:pro}.

The structure of the paper is as follows.
In Section~\ref{sec:review}, we first review the label shift
literature and some key results such as the efficient influence
function.
Section~\ref{sec:pro} presents the methodology and
Section~\ref{sec:theory} presents the corresponding theory,
respectively.
In Section~\ref{sec:discreteY}, we specifically discuss the discrete
label case which is of primary interest in the machine learning
community.
From Sections~\ref{sec:pro} to \ref{sec:discreteY}, the data available
to analysts are both labeled and unlabeled data, and the ML prediction
model available to analysts is $\wh\E_p(\cdot\mid \x)$, where
$\E_p$ denotes the expectation with respect to $\calP$.
Similar to the setting in \cite{angelopoulos2023prediction}, this ML
prediction model can be fitted from the labeled data $\calP$ or from
some external data that has the same distribution as the labeled
data.
Section~\ref{sec:simu} contains the simulation results and
Section~\ref{sec:data} contains two data applications, respectively. 
The paper is concluded with some discussions in Section~\ref{sec:disc}.
All the technical details are deferred in the Supplementary Material.

\section{Review: Label Shift and Efficient Influence Function (EIF)}\label{sec:review}

We start by briefly reviewing the label shift assumption and some
existing results under label shift.
Throughout the paper, we observe both source domain labeled data \eqref{eq:data:P} and target domain unlabeled data
\be\label{eq:data:QneqP}
\X_i, i=n+1,\dots,N, \mbox{ drawn as a random sample from population } \calQ,
\mbox{ where } \calP \neq \calQ.
\ee
We denote $\py$ and $\qy$ the density of $Y$ in $\calP$ and $\calQ$,
$\pxy$ and $\qxy$ the conditional density of $\X$ given $Y$ in $\calP$
and $\calQ$, respectively.

We emphasize that we allow the distributions in $\calP$ and $\calQ$ to
  be different.
In the literature, various discussions have explored different formats of distribution shifts \citep{quinonero2009dataset}, such as covariate shift, label shift, etc.
In this paper, we consider the label shift assumption,
\be\label{eq:ass}
\py(y)\neq\qy(y),\mbox{~~~and~~~}\pxy(\x,y)=\qxy(\x,y)\equiv\fxy(\x,y)\mbox{~for some~}\fxy.
\ee
This means that in both population $\calP$ and population $\calQ$, as
long as the label $Y$ is given, the covariate $\X$ follows the same
distribution. We use the notation $\fxy$ to emphasize that the $\X\mid Y$ distribution
is identical in the two distinct populations.
The distributions of $\calP$ and $\calQ$ differ because the marginal distributions of $Y$ are different.
Label shift has been shown to be reasonable and practical in the
anticausal learning setting \citep{scholkopf2012causal}, where $Y$
causes $\X$---examples include diseases causing symptoms or objects
causing sensory observations \citep{storkey2009training,
  zhang2013domain, du2014semi, iyer2014maximum, nguyen2016continuous,
  tasche2017fisher, kim2024retasa, lee2024doubly}.
It is also suitable in computer vision applications, such as
predicting object locations, directions, and human poses
\citep{martinez2017simple, yang2018position, guo2020ltf}.

In this paper, we consider the parameter of interest $\bt$
that satisfies the global characteristic
\be\label{eq:theta0}
\Eq\{\U(Y,\X,\bt)\}=\0,
\ee
where $\U(\cdot)$ is a pre-specified known function, and $\Eq$ denotes
the expectation with respect to $\calQ$.
The overarching goal of this paper is to investigate the efficient
estimation and inference of $\bt$, under the assumption in~\eqref{eq:ass}.

The definition of $\bt$ in \eqref{eq:theta0} is quite general.
It includes the minimizer of a given loss function $\ell(\cdot)$ in
the target population $\calQ$, such as $\argmin_\bt
\Eq\{\ell(Y,\X,\bt)\}$.
For illustration purposes, we sometimes also present the results for a
simpler scenario defined as
\be\label{eq:simpletheta}
\theta=\Eq\{s(Y,\X)\},
\ee
with $s(\cdot)$ a pre-specified known function. This corresponds to
$U(Y,\X,\theta)=s(Y,\X)-\theta$.

\subsection{The Robust Estimation Procedure}\label{sec:eifinlee}

In both machine learning and statistics, there has been significant
progress in developing methods for prediction and estimation under
label shift.
Here, we briefly review the robust estimation procedure from
\cite{lee2024doubly}, as it forms the basis of the new method proposed
in this paper.

When pooling the data from both populations together, to facilitate
the presentation, the population indicator $R$ is introduced in that
$R=1$ if the sample is from $\calP$ and $R=0$ if from $\calQ$, with
$r_i$ being its sample realization.
Further, the proportion of the sample from $\calP$ is defined as
$\pi=n/N$, and the density ratio of $Y$, given that the support
  of $\qy$ is contained in that of
  $\py$,
is defined as
\be\label{eq:define:rho}
\rho(y)=\qy(y)/\py(y),
\ee
which is not straightforward to estimate because of the absence of $Y$
observations from $\calQ$.
The key idea of \cite{lee2024doubly} is to construct a locally
efficient influence function $\bphi\eff^*$ by replacing $\rho(\cdot)$
in the efficient influence function (EIF) $\bphi\eff$ with an
arbitrary working model $\rho^*(\cdot)$.
This approach maintains the validity of the estimating
  function by ensuring it has mean zero at the true
  parameter, a property that holds even without requiring a 
    correct identification of $\rho(\cdot)$. 
Depending on how other nuisance components are handled, the estimators
proposed in \cite{lee2024doubly} enjoys some nice robustness
properties such as single flexibility and double flexibility.

For the simpler estimand in \eqref{eq:simpletheta}, the proposed
singly flexible estimator in \cite{lee2024doubly} is
\be\label{eq:est_general}
\wt\theta &=& \frac1{N-n}\sum_{i=n+1}^{N}
\wt w_i \wh\E_p\{ \wh a^*(Y)\rho^*(Y)\mid\x_i\}
\n\\&&
+\frac1n\sumi \rho^*(y_i)
\left[s(y_i,\x_i)-
\wt w_i \wh\E_p\{ \wh a^*(Y)\rho^*(Y)\mid\x_i\}
\right],
\ee
where $\wt w_i = [\wh \E_p\{
\rho^{*2}(Y) + \pi / (1 - \pi) \rho^*(Y) \mid \x_i\}]^{-1}$'s are the estimates of the sample weights,
$\wh a^*(y)$ can be obtained by solving the equation
\be\label{eq:eq_general}
s(y,\x_i)
&=&\sumI \wt w_i\wh\E_p\{a(Y)\rho^*(Y)\mid\x_i\}\frac{r_i \wt K_l(y-y_i)}{\sumJ r_j \wt K_l(y-y_j)},
\ee
where $\sumI \ell(\x_i)\frac{r_i \wt K_l(y-y_i)}{\sumJ r_j \wt
  K_l(y-y_j)}$ is the approximation of the conditional expectation
$\E\{\ell(\X)\mid y\}$, $\wt K(\cdot)$ is the kernel function and $l$
is the bandwidth. This equation is known as a Fredholm equation of the
first kind, for which a detailed discussion and methods to solve
  it are
 in \cite{lee2024doubly}.
Note that, the EIF for estimating $\theta$ is
\bse
\phi\eff
&=&\frac{r}{\pi}\rho(y)\left[s(y,\x)-w(\x) \Ep\{ a(Y)\rho(Y)\mid\x\}\right]
+\frac{1-r}{1-\pi}\left[w(\x) \Ep\{ a(Y)\rho(Y)\mid\x\}-\theta\right],
\ese
where
$w(\x) = \left[ \Ep\{\rho^2(Y)+\pi/(1-\pi)\rho(Y)\mid\x\}\right]^{-1}$,
and $a(y)$ satisfies
\bse
\E\left[w(\X)\Ep\{ a(Y)\rho(Y)\mid\X\}\mid y\right]=s(y,\x).
\ese

For convenience of later use, we summarize this robust
estimation procedure below in Algorithm~\ref{alg:general}, and call it
\bse
\textsc{Algorithm}\ [\theta, \rho^*(y)],
\ese
where the first argument $\theta$ indicates the estimand and the
second argument $\rho^*(y)$ indicates how the density ratio $\rho(y)$
is implemented in the algorithm.

\begin{algorithm}[tbp]
\caption{Algorithm [$\theta=\Eq\{s(Y,\X)\}$, $\rho^*(y)$]}\label{alg:general}
\begin{algorithmic}
  \State \textbf{Data Input}: data from $\calP$: $(r_i=1, y_i, \x_i)$, $i=1,\dots,n$, data from $\calQ$: $(r_j=0, \x_j)$, $j=n+1,\dots,N$, and $\pi=n/N$.
  \State \textbf{Algorithm/Model Input}: the ML prediction model $\wh \E_p(\cdot\mid \x)$, and the density ratio model $\rho^*(y)$.
  \State \textbf{Output}: $\wt\theta$, as defined in \eqref{eq:est_general}.
  \State \textbf{Do}:
  \State (a) compute $\wt w_i=[\wh\E_p\{\rho^{*2}(Y)+\pi/(1-\pi)\rho^*(Y)\mid\x_i\}]^{-1}$ for $i=1,\dots,N$;
  \State (b) obtain $\wh a^{*}(y)$ by solving \eqref{eq:eq_general};
  \State (c) construct $\wt\theta$ as in \eqref{eq:est_general}.
\end{algorithmic}
\end{algorithm}

\subsection{The Limitation and Our Novel Contributions}\label{sec:limit}

An obvious limitation of \cite{lee2024doubly} is that the proposed estimators are not efficient.
The performance of their estimators heavily depends on how lucky they
are in choosing $\rho^*(\cdot)$: if the working model
$\rho^*(\cdot)$ is unfortunately very badly
chosen, their estimators may have substantial standard error and poor
inference performance.

In contrast, we propose an efficient estimator with the smallest
variability among all regular asymptotically linear estimators
\citep{bickel1993efficient, tsiatis2006semiparametric}.
A key step in achieving efficiency is estimating $\rho(\cdot)$, a
step omitted in \cite{lee2024doubly}.
Notably, although $\rho(y)$ is a function of $y$ hence describes
  \emph{local} features of $\calQ$,
estimating $\rho(y)$ can almost be reframed as a problem of
estimating a \emph{global} feature in $\calQ$, using a specially defined
$\U(\cdot)$ function in \eqref{eq:theta0}.
While technical challenges must be addressed, this insight is central
to developing our efficient estimator for the general parameter
$\bt$. The details will be presented in Section~\ref{sec:pro}.


\section{Proposed Methodology}\label{sec:pro}

Note that $\rho(y)$ is defined as the density ratio $\rho(y) = \qy(y) / \py(y)$.
Between $\py(y)$ and $\qy(y)$, estimating $\py(y)$ is relatively
straightforward since $Y$-data are available in the population
$\calP$. However, estimating $\qy(y)$ is challenging due to the
absence of observations.
In Section~\ref{sec:rho}, we first introduce an initial consistent
estimator for $\rho(y)$, which is already sufficient for the efficient
estimation of $\bt$ proposed in Section~\ref{sec:general-eff}.
Additionally, an efficient estimation of $\rho(y)$ is also feasible
and is presented in Section~\ref{sec:rho-eff}.
The progression in estimating $\rho(y)$---from the heuristic guess in
\cite{lee2024doubly}, to the consistent estimation in
Section~\ref{sec:rho}, and finally to the efficient estimation in
Section~\ref{sec:rho-eff}, is notably achieved by repeating the same procedure, hence
can be viewed as a self-driven evolutionary process.


\subsection{An Initial Step for Consistently Estimating $\rho(y)$}\label{sec:rho}

For estimating $q_Y(y)$, without direct observations, we
must infer information
about $\qy(y)$ indirectly through the available $\X$
observations from
$\calQ$ and the link between $\calP$ and $\calQ$. Therefore,
the core difficulty in estimating $\rho(y)$ lies
in constructing a consistent estimator of $\qy(y)$ that can
overcome the absence of the outcome in $\calQ$
by leveraging the information in the covariates from $\calQ$,
the data from $\calP$, and the distributional link between the two
populations.

Consider the estimation of $\qy(y)$ at an arbitrary
point $y = y_0$ in the support of $Y$.
The quantity $\qy(y_0)$ is essentially
a local feature of $\calQ$, whereas the method reviewed in
\textsc{Algorithm [$\theta, \rho^*(y)$]}
applies only to a quantity $\theta$ that represents
a global feature of $\calQ$.
To bridge this gap, we approximate $\qy(y_0)$
by a global feature
and thereby facilitate its estimation.
Specifically,
we can approximate $\qy(y_0)$ by
\bse
\delta_0 \equiv \Eq\{K_h(Y -
y_0)\},
\ese
where $K_h(y) = K(y/h) / h$, with $K(\cdot)$ being a kernel
function and $h$ a bandwidth.
Now, we can view $\delta_0$ as an unknown parameter in $\calQ$ defined in \eqref{eq:simpletheta} corresponding to $s(Y,\X)=K_h(Y - y_0)$.
This formulation allows for utilization of the EIF for estimating $\delta_0$, with its detailed derivation presented in Supplementary Material \ref{supp:sec:deltaEIF}.
Thus, one can apply
the \textsc{Algorithm [$\delta_0, \rho^*(y)$]}
to obtain the estimator
\bse
\wt\delta_0 &=& \frac1{N-n}\sum_{i=n+1}^{N}
\wt w_i \wh\E_p\{ \wh a^*(Y)\rho^*(Y)\mid\x_i\}
\\&&
+\frac1n\sumi \rho^*(y_i)
\left[K_h(y_i-y_0)-
\wt w_i \wh\E_p\{ \wh a^*(Y)\rho^*(Y)\mid\x_i\}
\right],
\ese
where $\wt w_i$ is the same as that in \eqref{eq:est_general}, $\wh a^*(y)$ is also the same as the solver of \eqref{eq:eq_general} except for replacing $s(y,\x)$ by $K_h(y-y_0)$.
Consequently, the estimator of $\rho(y_0)$ is obtained as
\be\label{eq:est_rho_consistent}
\wt\rho(y_0)=\frac{\wt\delta_0}{\wh p_Y(y_0)}\quad\text{where}\quad \wh p_Y(y_0)=n^{-1}\sumi K_h(y_i-y_0).
\ee
Note that we can estimate  $\rho(y)$ at any $y=y_0$
  value above, hence we can estimate the entire function $\rho(\cdot)$,
  i.e., we can obtain $\wt\rho(\cdot)$.
  In practice, we may preselect a set of points
  $\{t_1,\dots,t_{M_n}\}$ on the support and implement the estimator
  $\wt\rho(y_0)$ at each $y_0=t_k$ for $k=1,\dots,M_n$, then construct
  a smoothing curve that interpolates
  $\{\wt\rho(t_1),\dots,\wt\rho(t_{M_n})\}$ to establish
  $\wt\rho(\cdot)$. Here $M_n$ can be set to $[n^{1/4}]$ to ensure
    sufficient precision.
\begin{remark}\label{remark:workingmodel}
  To ensure consistency in presentation throughout the paper, even for
  the purpose of estimating $\rho(y)$, we assume here that an ML
  prediction model $\wh \E_p(\cdot \mid \x)$ is available.
  However, it is important to highlight that a correctly specified
  model of $\Ep(\cdot \mid \x)$ is NOT required for the purpose of
  estimating $\rho(y)$ in the above procedure.
  This is a key attribute of the locally efficient influence function,
  as discovered in \cite{lee2024doubly} and referred to as ``double
  flexibility''.
  This characteristic distinguishes our proposed method here from
  others in the literature, such as \cite{kim2024retasa} which relies
  on a correctly specified model of $\Ep(\cdot \mid \x)$ for
  estimating $\rho(y)$.
\end{remark}


\subsection{Efficiently Estimating $\rho(y)$}\label{sec:rho-eff}

Equipped with $\wt\rho(y)$, one can further refine the estimation for
$\rho(y)$ by first applying the \textsc{Algorithm [$\delta_0,
  \wt\rho(y)$]} to obtain $\wh \delta_0$, where the input model for the
density ratio is replaced with the estimated $\wt\rho(y)$ in
Section~\ref{sec:rho} in lieu of $\rho^*(y)$, then formulating the
refined estimator as
\be\label{eq:wtrho}
\wh\rho(y_0)=\frac{\wh\delta_0}{\wh p_Y(y_0)}\quad\text{where}\quad \wh p_Y(y_0)=n^{-1}\sumi K_h(y_i-y_0),
\ee
and
\bse
\wh\delta_0&=& \frac1{N-n}\sum_{i=n+1}^N \wh w_i\wh\E_p\{\wh a(Y)\wt\rho(Y)\mid\x_i\}
\\&&
+ \frac1n\sumi\wt\rho(y_i)\left[K_h(y_i-y_0)-\wh w_i\wh\E_p\{\wh a(Y)\wt\rho(Y)\mid\x_i\}\right],
\ese
where $\wh w_i = [\wh \E_p\{ \wt\rho^2(Y) + \pi / (1 - \pi) \wt\rho(Y) \mid \x_i\}]^{-1}$, and $\wh a(y)$ represents the solution by solving the equation \eqref{eq:eq_general} in the \textsc{Algorithm [$\delta_0, \wt\rho(y)$]}.


This refinement procedure further reduces the variability of the $\rho(y)$ estimator.
In fact, as we will demonstrate in Section \ref{sec:theory}, the
resulting estimator $\wh\rho(y)$ achieves the minimum variance at a
given kernel function and bandwidth.
Thus, we complete a self-improvement process by iteratively applying
our method, starting from a likely  inconsistent initial guess
$\rho^*(y)$, progressing to
a consistent estimate $\wt\rho(y)$, and ultimately achieving
an efficient estimator $\wh\rho(y)$.

\begin{remark}\label{remark:onestep}
The self-driven evolutionary process of iteratively applying the same
algorithm to achieve the efficient estimation of $\rho(y)$
given fixed kernel function and bandwidth, as described in
Sections~\ref{sec:rho} and~\ref{sec:rho-eff}, is fundamentally
different from the classical one-step estimation methods in
semiparametric statistics and targeted learning \citep{bickel1975one,
  bickel1993efficient, van2011targeted}. 
Though we estimate $\rho(y)$ via estimating $\py(y_0)$ and $\qy(y_0)$
at each support point $y_0$, our goal here is to estimate the entire
function $\rho(y)$ instead of a parameter with fixed dimensionality.
More importantly, a key innovation in our approach is the
approximation of $\qy(y_0)$ by $\delta_0$, which can then be estimated
via implementing the locally efficient influence function.
This approximation strategy is central not only in
Section~\ref{sec:rho}, but also in the efficient estimation procedure
of Section~\ref{sec:rho-eff}, where $\rho(y)$ is also treated as a
nuisance function and estimated by its consistent estimator
$\wt\rho(y)$. 
In contrast, classical one-step estimation does not rely on such
approximations: it typically begins with an initial estimator that
does not use the efficient influence function, and then achieves
efficiency by updating an augmentation term. 
Our approach, while technically more involved, introduces a novel and
necessary idea that enables efficient estimation of the general
parameter $\bt$. 
\end{remark}

\subsection{Efficiently Estimating $\theta$ }\label{sec:general-eff}

We now return to the primary objective of our work:
to devise an efficient estimator of the general parameter $\bt$
defined in \eqref{eq:theta0}.
Given the consistent estimate of $\rho(y)$, $\wt\rho(y)$, presented in Section~\ref{sec:rho},
the procedure is overall similar to the construction of
$\wh\delta_0$ in Section~\ref{sec:rho-eff}.
For estimating $\bt$, the EIF \citep{lee2024doubly} is
\be\label{eq:eff}
 \bphi\eff(r,ry,\x,\bt)
&=&\A(\bt) \left( \frac{r}{\pi}\rho(y)\left[\U(y,\x,\bt)
-w(\x)\Ep\{\U(Y,\x,\bt)\rho^2(Y)+\a(Y)\rho(Y)\mid\x\}\right]\right.\n\\
&&\left.+\frac{1-r}{1-\pi}w(\x)\Ep\{\U(Y,\x,\bt)\rho^2(Y)+\a(Y)\rho(Y)\mid\x\}\right),
\ee
where $\A(\bt)\equiv[\Eq\{\partial\U(Y,\X,\bt)/\partial\bt\trans\}]^{-1}$,
$w(\x) = \left[\Ep\{\rho^2(Y)+\pi/(1-\pi)\rho(Y)\mid\x\}\right]^{-1}$,
and $\a(y)$ needs to satisfy
\bse
\E\left[w(\X)\Ep\{\U(Y,\X,\bt)\rho^2(Y)+\a(Y)\rho(Y)\mid\X\}
\mid y\right]
=\E\{\U(y,\X,\bt)\mid y\}.
\ese

To be more specific, given the consistent estimate $\wt\rho(y)$ and
the ML prediction model $\wh\E_p(\cdot\mid \x)$, we first compute the
estimate for the weight function
$\wh w_i = [\wh
\E_p\{\wt \rho^2(Y) + \pi / (1 - \pi) \wt \rho(Y) \mid \x_i\}]^{-1}$
at each observation $i=1, \dots, N$.
Then, we obtain the function $\wh\a(y)$ by solving the integral equation
\bse
&&\sum_{i=1}^N \wh w_i\wh\E_p\{\a(Y)\wt\rho(Y)\mid\x_i\}r_i\wt K_l(y-y_i)\\
&=&\sum_{i=1}^N\left[\U(y,\x_i,\bt)-\wh w_i\wh\E_p\{\U(Y,\x_i,\bt)\wt\rho^2(Y)\mid\x_i\}\right]r_i\wt K_l(y-y_i)
\ese
with respect to $\a(y)$, where $\wt K_l(y-y_i) =\wt K\{(y-y_i)/l\}/l$,
$\wt K(\cdot)$ is a kernel function, and $l$ is a bandwidth, chosen to
approximate $\E(\cdot\mid y)$ by smoothing over the neighborhood of
$y$.

Afterwards, we calculate
\be\label{eq:predictU}
\wh\b(\x_i) =
\wh w_i\wh\E_p\{\U(Y,\x_i,\bt)\wt\rho^2(Y) + \wh\a(Y)\wt\rho(Y) \mid \x_i\},
\ee
for each sample $i=1, \dots, N$.
Finally, we obtain $\wh\bt$ by solving the following estimating equation
\be\label{eq:solve_bt}
\underbrace{\frac{1}{N-n}\sum_{i=n+1}^N \wh\b(\x_i)}_{\substack{\text{Direct estimating equation using}\\ \text{predictions on unlabeled data}}} \quad + \quad
\underbrace{\frac1n\sum_{i=1}^n\left[\wt\rho(y_i)\{\U(y_i,\x_i,\bt) - \wh\b(\x_i)\}\right]}_{\substack{\text{Rectifier: measure the bias of using}\\ \text{predictions on labeled data}}} &=& \0,
\ee
with respect to $\bt$.
The summary of the implementation can be found in the
Algorithm~\ref{alg:general:theta0} below.
It can be seen that for solving $\wh\bt$, the term $\wh \b(\x_i)$
serves as the predicted value of the estimating equation $\U$, say,
$\wh \U(y_i,\x_i,\bt)$.

\begin{algorithm}[tbp]
\caption{Algorithm for efficiently estimating $\bt$ defined in \eqref{eq:theta0}}\label{alg:general:theta0}
\begin{algorithmic}
  \State \textbf{Data Input}: data from $\calP$: $(r_i=1, y_i, \x_i)$, $i=1,\dots,n$, data from $\calQ$: $(r_j=0, \x_j)$, $j=n+1,\dots,N$, and $\pi=n/N$.
  \State \textbf{Algorithm/Model Input}: the ML prediction model $\wh \E_p(\cdot\mid \x)$, and the density ratio model $\rho^*(y)$.
  \State \textbf{Output}: $\wh\bt$, via solving \eqref{eq:solve_bt}.
  \State \textbf{Do}:
  \State (a) follow \textsc{Algorithm [$\delta_0, \rho^*(y)$]} to compute $\wt\delta_0$ and then compute $\wt\rho(y)$ in \eqref{eq:est_rho_consistent};
    \State (b) follow \textsc{Algorithm [$\bt, \wt\rho(y)$]} to compute $\wh\bt$, via solving \eqref{eq:solve_bt}.
\end{algorithmic}
\end{algorithm}

To conclude this section, we elaborate a little  more on the efficient
estimator  for the simplest case $\theta=\E_q(Y)$, where the efficient estimator can be simplified as
\be\label{eq:est_general:illustration}
\wh\theta &=& \underbrace{\frac1{N-n}\sum_{i=n+1}^{N} \wh y_i}_{\substack{\text{Direct estimation}\\ \text{using predictions}}}
\quad +
\underbrace{\frac1n\sumi \wt\rho(y_i)(y_i-\wh y_i)}_{\substack{\text{Rectifier: measure the bias of using}\\ \text{predictions on labeled data}}},
\ee
with $\wh y_i = \wh w_i \wh\E_p\{ \wh a(Y)\wt\rho(Y)\mid\x_i\}$.
Clearly, our proposed efficient estimator has the similar
  appearance as the existing PPI-type estimator in the absence of
distribution shifts, e.g., in \cite{angelopoulos2023prediction}.
However, due to the presence of label shift and the goal of achieving
efficient estimation, our estimator incorporates several key
differences compared to the original PPI estimator.
First, to ensure efficient estimation, the density ratio model
$\rho(y)$ plays a crucial role in the ``rectifier'' component.
Accurate estimation of $\rho(y)$ is essential in the presence of label
shift, whereas in the original PPI framework, $\rho(y)=1$ by
assumption.
Second, the construction of the predictive values $\wh y_i$'s is more
sophisticated.
Unlike the simple conditional mean estimates obtained from an ML
prediction model, our approach defines the predictive values $\wh
y=\wh\b(\x)$ as in \eqref{eq:predictU}.
This structure ensures that $\wh\E(\wh Y\mid
y) =\wh\E\{\wh\b(\X)\mid y\} = y$ where $\wh\E$ is the approximation to the expectation of $\X\mid Y$, guarantees the consistency of the direct estimation term in
\eqref{eq:est_general:illustration} for $\Eq(Y)$. Furthermore, $\wh 
y=\wh\b(\x)$ is constructed to make the estimator in
\eqref{eq:est_general:illustration} efficient.
In contrast, in the original PPI, the direct estimation
  term in \eqref{eq:PPI} can be biased. Even when it is unbiased,
  the estimator is not efficient in the presence of label shift.


\section{Asymptotic Theory}\label{sec:theory}

\subsection{Asymptotic Properties of the Estimator in Section~\ref{sec:rho}}\label{sec:rho:theory}

As we outlined in Remark~\ref{remark:workingmodel}, a key feature of
the proposed estimator $\wt\rho(y)$ in Section~\ref{sec:rho} is that
its implementation does NOT require a correctly specified model of
$\Ep(\cdot \mid \x)$ to consistently estimate
  $\rho(y)$.
To present our theory reflecting this flexibility in the choice of $\Ep(\cdot \mid \x)$, we use superscript $^\dag$ throughout this Section~\ref{sec:rho:theory}.
Specifically, we use $\pyx^\dag(y,\x)$ and
$\Ep^\dag(\cdot\mid \x)$ to denote the chosen
conditional distribution function and conditional expectation with
respect to the model of $Y$ given $\X$ in the labeled
population $\calP$ that may or may not be correct.

To facilitate the presentation of the theoretical properties, we introduce some more notations. Let us denote 
\bse
u^{*\dag}(t,y)
&\equiv&\py(y)\int\frac{\rho^*(t)\pyx^\dag(t,\x)}
{\Ep^\dag\{\rho^{*2}(Y)+\pi/(1-\pi)\rho^*(Y)\mid\x\}}\fxy(\x,y)d\x,\\
\calI^{*\dag}(a)(y)
&\equiv&\py(y)\E\left[\frac{\Ep^\dag\{a(Y)\rho^*(Y)\mid\X\}}
{\Ep^\dag\{\rho^{*2}(Y)+\pi/(1-\pi)\rho^*(Y)\mid\X\}}\mid y\right]
=\int a(t)u^{*\dag}(t,y)dt,\\
v_h(y)
&\equiv&\py(y)K_h(y-y_0),
\ese
and let $a_h^{*\dag}(y)$ satisfy
$\calI^{*\dag}(a_h^{*\dag})(y)=v_h(y)$.
For a function $a(y,\x)\in\mathbb{R}$, let $\|a(y,\x)\|_p=[\int \{a(y,\x)\}^pdyd\x]^{1/p}~(1\leq p<\infty)$ be the $L_p$-norm of $a(y,\x)$, and let $\|a(y,\x)\|_\infty=\sup_{y,\x}|a(y,\x)|$ be the sup-norm of $a(y,\x)$. In addition, for a vector-valued function $\a(y,\x)=\{a_1(y,\x),\dots,a_d(y,\x)\}\trans\in\mathbb{R}^d$, let $\|\a(y,\x)\|_\infty=\max_{k=1,\dots,d}\|a_k(y,\x)\|_\infty$.

We require the following regularity conditions to develop our theory.
\begin{longlist}
\renewcommand{\labellonglist}{\thelonglist}
\renewcommand{\thelonglist}{(A\arabic{longlist})}
    \item\label{con:rho}
    $\rho^*(y)$ is bounded away from zero on the support of $\py(y)$,
    and has a bounded $k$th derivative for
    $k=1,\dots,m$.
    \item\label{con:pyx}
    $\pyx^\dag(y,\x)$ is complete, i.e.,
    $\Ep^\dag\{g(Y)\mid\x\}=\int g(y)\pyx^\dag(y,\x)dy=0$ for
    all $\x$ implies $g(y)=0$ almost
    surely. The function $\rho^*(\cdot)\pyx^\dag(\cdot,\x)$
    is square integrable.
    \item\label{con:u}
    $u^{*\dag}(t,y)$ is square integrable, and has bounded $k$th
    derivatives with respect to $t$ and $y$ for $k=1,\dots,m$.
    \item\label{con:density}
    The supports of $\fxy(\x,y),\py(y),\rho(y)$ are compact. $\rho(y)$ is bounded, and $p(y)$
    and $q(y)$ have bounded $k$th derivatives for $k=1,\dots,m$.
    \item\label{con:kernel}
    $K(\cdot)$ and $\wt K(\cdot)$ have compact support $[-1,1]$, are of order $m$, bounded,
    and
    have bounded $k$th derivatives for $k=1,\dots,m$.
    \item\label{con:bandwidth}
    $h\to0$, $\{n\min(n,N-n)\}^{1/2}h\to\infty$, $nl^{2m}\to 0$, $nl^2\to\infty$, and $nh^{-2m-1}l^{2m}\to0$ as $n\to\infty$.
\end{longlist}
The requirement on the bandwidth $l$ in Condition \ref{con:bandwidth}
regulates the Nadaraya-Watson estimator of $\E(\cdot\mid y)$ based on
$\wt K_l(\cdot)$ to converge at rate
$o_p(n^{-1/4})$. To implement our
estimator in practice, the bandwidths can be chosen as follows:
suppose we adopt a kernel function $K(\cdot)$ that is of order $m=2$,
Gaussian for instance, then the bandwidth $l$ needs to be of order
between $n^{-1/2}$ and $n^{-1/4}$ according to Condition
\ref{con:bandwidth}. If we choose $l=C_1n^{-1/3}$ for some constant
$C_1$, then subsequently the other bandwidth $h$ needs to be of order
greater than $n^{-1/15}$ and still converge to zero. In our numerical
studies, we chose $h=C_2n^{-1/16}$ for some constant $C_2$.

Below, we present three results, corresponding to the behavior of the
operator $\calI^{*\dag}$, the asymptotic normal distribution of the
estimator $\wt\delta_0$, and the convergence rate of
the estimator $\wt\rho(y)$, respectively.
Their proofs are contained in Supplementary Material~\ref{supp:sec:prooflemma1},
\ref{supp:sec:proofthm1}, \ref{supp:sec:proofthm2}, respectively.

\begin{lemma}\label{lem:i}
Under Conditions \ref{con:rho}-\ref{con:density},
\begin{longlist}
    \item the linear operator $\calI^{*\dag}: L^2(\bbR)\to L^2(\bbR)$ is invertible,
    \item there exist constants $C_1, C_2$ such that $0<C_1, C_2<\infty$, $\|\calI^{*\dag}(a)\|_2\leq C_1\|a\|_2$, and $\|\calI^{*\dag-1}(a)\|_2\leq C_2\|a\|_2$ for all $a(y)\in L^2(\bbR)$.
\end{longlist}
\end{lemma}

\begin{theorem}\label{th:theta}
Assume Conditions \ref{con:rho}-\ref{con:bandwidth}. 
For any choice of $\rho^*(y)$
and $\Ep^\dag(\cdot\mid\x)$, 
at any $y_0$,
$\var(\wt\delta_0)^{-1/2}\{\wt\delta_0-\bias(\wt\delta_0)-\qy(y_0)\}$
converges in distribution to the standard normal distribution, where
\bse
\bias(\wt\delta_0)&=&\qy^{(m)}(y_0)\frac{\int t^mK(t)dt}{m!}h^m +
O(h^{m+1})+o(n^{-1/2}h^{-1/2}),\\
\var(\wt\delta_0)&=&n^{-1}\Ep\left\{\left(\rho(Y)\left[K_h(Y-y_0)
-\frac{\Ep^\dag\{a_h^{*\dag}(Y)\rho^*(Y)\mid\X\}}
{\Ep^\dag\{\rho^{*2}(Y)+\pi/(1-\pi)\rho^*(Y)\mid\X\}}\right]\right)^2\right\}\\
&&+(N-n)^{-1}\Eq\left(\left[\frac{\Ep^\dag\{a_h^{*\dag}(Y)\rho^*(Y)\mid\X\}}
{\Ep^\dag\{\rho^{*2}(Y)+\pi/(1-\pi)\rho^*(Y)\mid\X\}}-\delta_0\right]^2\right)\\
&&+o[\{\min(n,N-n)\}^{-1/2}n^{-1/2}h^{-1}].
\ese
\end{theorem}

Theorem \ref{th:theta} guarantees that our proposed
  estimator $\wt\delta_0$ is consistent for $q_Y(y_0)$ regardless
  the chosen models for $\rho(y)$ and $\Ep(\cdot\mid\x)$ are
  correct or not. According to the bias and standard error of $\wt\delta_0$
  presented in Theorem \ref{th:theta}, $\wt\delta_0$ converges to
  $q_Y(y_0)$ as long as the bandwidth $h$ is chosen to satisfy the
  regularity conditions in Condition \ref{con:bandwidth}. Since
  this result holds under arbitrary models $\rho^*(y)$ and
  $\Ep^\dag(\cdot\mid\x)$, we are allowed to freely choose the initial
  guess for $\rho(y)$ and the prediction model for
  $\Ep(\cdot\mid\x)$.

Based on the above point-wise convergence behavior of the estimator of
$\qy(y_0)$, we further elaborate the uniform convergence rate of
$\wt\rho(\cdot)$. For simplicity,
we consider the function $\wt\rho(\cdot)$ constructed
as a piecewise linear interpolation of $\{\wt\rho(t_1),\dots,\wt\rho(t_{M_n})\}$.
In short, the estimator converges uniformly to
the true $\rho(y)$ under a few additional regularity conditions.
Such convergence allows us to successfully
implement the EIF for estimating a general parameter~$\bt$.

\begin{theorem}\label{th:rhomax}
Assume Conditions \ref{con:rho}-\ref{con:bandwidth}. In addition,
assume $\rho(y)$ is Lipschitz continuous. Then for any choice of $\rho^*(y)$ and $\Ep^\dag(\cdot\mid\x)$, 
\bse
\|\wt\rho(y)-\rho(y)\|_\infty=O(h^m)+O_p[\{\min(n,N-n)h\}^{-1/2}\log n]=o_p(1).
\ese
\end{theorem}

\subsection{Asymptotic Properties of the Estimator in
  Section~\ref{sec:rho-eff}}\label{sec:rho-eff:theory}

To state the theoretical properties of $\wh\delta_0$ and $\wh\rho(y)$ studied in Section~\ref{sec:rho-eff}, we
introduce some more notations.
Define
\bse
\wt u(t,y)
&\equiv&\py(y)\int\frac{\wt\rho(t)\pyx(t,\x)}
{\Ep\{\wt\rho^2(Y)+\pi/(1-\pi)\wt\rho(Y)\mid\x\}}\fxy(\x,y)d\x,\\
\wt\calI(a)(y)
&\equiv&\py(y)\E\left[\frac{\Ep\{a(Y)\wt\rho(Y)\mid\X\}}
{\Ep\{\wt\rho^2(Y)+\pi/(1-\pi)\wt\rho(Y)\mid\X\}}\mid y\right]
=\int a(t)\wt u(t,y)dt,
\ese
and let $\wt a_h(y)$ satisfy $\wt\calI(\wt a_h)(y)=v_h(y)$.
We also need the following regularity conditions to develop the theory.
\begin{longlist}
\renewcommand{\labellonglist}{\thelonglist}
\renewcommand{\thelonglist}{(A\arabic{longlist}')}
    \item\label{con:rho'}
    $\wt\rho(y)$ is bounded away from zero on the support of $\py(y)$, and has a bounded $k$th derivative for $k=1,\dots,m$.
    \item\label{con:pyx'}
    $\pyx(y,\x)$ is complete, i.e., $\Ep\{g(Y)\mid\x\}=\int g(y)\pyx(y,\x)dy=0$ for all $\x$ implies $g(y)=0$ almost surely. The function $\wt\rho(\cdot)\pyx(\cdot,\x)$ is square integrable.
    \item\label{con:u'}
    $\wt u(t,y)$ is square integrable, and has bounded $k$th derivatives with respect to $t$ and $y$ for $k=1,\dots,m$.
\end{longlist}

The following result concerns the asymptotic normality as well as the estimation efficiency of $\wh\delta_0$, with the proof contained in Supplementary Material~\ref{supp:sec:proofthm3}.

\begin{theorem}\label{th:thetaeff}
Assume Conditions \ref{con:rho'}-\ref{con:u'}, \ref{con:density}-\ref{con:bandwidth}.
If $\wt\rho(\cdot)$ satisfies $\sup_y|\wt\rho(y)-\rho(y)|=o_p(1)$ and
$\wh\E_p(\cdot\mid \x)$ satisfies
$|\wh\E_p\{a(Y)\mid\x\}-\Ep\{a(Y)\mid\x\}|=o_p(n^{-1/4}\|a\|_2)$ for
every square integrable $a(y)$ and $\x$,
then for any $y_0$,
$\var(\wh\delta_0)^{-1/2}\{\wh\delta_0-\bias(\wh\delta_0)-\qy(y_0)\}$ converges in distribution to the standard normal distribution, where
\bse
\bias(\wh\delta_0)&=&\qy^{(m)}(y_0)\frac{\int t^mK(t)dt}{m!}h^m +
O(h^{m+1})+o(n^{-1/2}h^{-1/2}),\\
\var(\wh\delta_0)&=&n^{-1}\Ep\left\{\left(\rho(Y)\left[K_h(Y-y_0)-
\frac{\Ep\{a_h(Y)\rho(Y)\mid\X\}}
{\Ep\{\rho^2(Y)+\pi/(1-\pi)\rho(Y)\mid\X\}}\right]\right)^2\right\}\\
&&+(N-n)^{-1}\Eq\left(\left[\frac{\Ep\{a_h(Y)\rho(Y)\mid\X\}}
{\Ep\{\rho^2(Y)+\pi/(1-\pi)\rho(Y)\mid\X\}}-\delta_0\right]^2\right)\\
&&+o[\{\min(n,N-n)\}^{-1/2}n^{-1/2}h^{-1}].
\ese
\end{theorem}

Theorem~\ref{th:thetaeff} indicates that using suitable
nuisance estimators for $\rho$ and $\Ep$ and bandwidth $h$ chosen as
Condition~\ref{con:bandwidth},
$\bias(\wh\delta_0)$ vanishes and $\var(\wh\delta_0)$ achieves the
semiparametric efficiency bound for estimating $\Eq\{K_h(Y-y_0)\}$,
where the corresponding EIF is derived in
Supplementary Material~\ref{supp:sec:deltaEIF}.
Hence, at fixed choices of $K(\cdot)$ and $h$, $\wh\delta_0$ is the
semiparametrically efficient estimator of $q_Y(y_0)$.
An interesting observation is that for the nuisance estimator $\wt\rho$,
uniform consistency is already sufficient to guarantee the efficiency of our
proposed estimator $\wh\delta_0$,
and no specific convergence order is imposed on $\wt\rho$.

In terms of estimating $\rho(y_0)=\qy(y_0)/\py(y_0)$ at an arbitrary
$y_0$, we can construct $\wh\rho(y_0)=\wh q_Y(y_0)/\wh
p_Y(y_0)$ where $\wh p_Y(y_0)$ is the standard Nadaraya-Watson kernel
estimator.
Because $\wh p_Y(y_0)$ and $\wh q_Y(y_0)$ are both efficient at the
given kernel function $K(\cdot)$ and bandwidth $h$, by the
delta-method, we can see that $\wh\rho(y_0)$ is also efficient.
We summarize this result in Corollary \ref{cor:effrho} below and omit its proof.

  \begin{corollary}\label{cor:effrho}
    At any $y_0$,
    the estimator $\wh\rho(y_0)=\wh q_Y(y_0)/\wh p_Y(y_0)$, where $\wh p_Y(y_0)$ is the Nadaraya-Watson
    estimator and $\wh q_Y(y_0)$ is given in Theorem \ref{th:thetaeff},
    is efficient given the kernel function $K(\cdot)$ and the bandwidth $h$.
\end{corollary}

We further present the uniform convergence result of
$\wh\rho(y)$. Since the bias and the variance of $\wh\rho(y_0)$ are of
the same order as those of $\wt\rho(y_0)$ as presented in Theorems
\ref{th:theta} and \ref{th:thetaeff}, $\|\wh\rho(y)-\rho(y)\|_\infty$
is also of the same order as $\|\wt\rho(y)-\rho(y)\|_\infty$ presented
in Theorem \ref{th:rhomax}. We state this result in Corollary
\ref{cor:whrhomax} below and omit its proof.

\begin{corollary}\label{cor:whrhomax}
  Assume the same conditions as in Theorem \ref{th:thetaeff}.
In addition, assume  $\rho(y)$ is Lipschitz continuous. Then
\bse
\|\wh\rho(
  \cdot)-\rho(\cdot)\|_\infty=O(h^m)+O_p[\{\min(n,N-n)h\}^{-1/2}\log n]=o_p(1).
\ese
\end{corollary}

\subsection{Asymptotic Properties of the Estimator in Section~\ref{sec:general-eff}}\label{sec:general-eff:theory}

For the estimator $\wh\bt$, its asymptotic normality and semiparametric efficiency can be found below, with the proof contained in Supplementary Material~\ref{supp:sec:proofthm4}.
We first list some additional regularity conditions.

\begin{longlist}
\renewcommand{\labellonglist}{\thelonglist}
\renewcommand{\thelonglist}{(B\arabic{longlist})}
    \item\label{con:0existence}
    $\Eq\{\partial\U(Y,\X,\bt)/\partial\bt\trans\}$ and $\E\{\partial\bphi\eff(R,RY,\X,\bt)/\partial\bt\trans\}$
    are invertible,
    $\bt\in\boldsymbol{\Theta}$ where $\boldsymbol{\Theta}$ is compact,
    and $\E\{\sup_{\bt\in\boldsymbol{\Theta}}\|\bphi\eff(R,RY,\X,\bt)\|_2\}<\infty$.
    $\U(y,\x,\bt)$ is twice differentiable with respect to $y$
    and its derivative is bounded.
    \item\label{con:3bounded2}
    The function $u(t,y)$ is bounded
    and has bounded derivatives with respect to $t$ and $y$ on its support.
    $\a(y)$ in \eqref{eq:eff} is bounded.
    \item\label{con:6bandwidth2}
    The bandwidth $l$ satisfies $n(\log n)^{-4}l^2\to\infty$ and
    $N^2n^{-1}l^4\to 0$.
\end{longlist}

\begin{theorem}\label{th:general-eff}
Assume Conditions \ref{con:rho'}, \ref{con:pyx'}, \ref{con:density}, \ref{con:kernel}, \ref{con:0existence}-\ref{con:6bandwidth2}. Suppose $\wt\rho(\cdot)$ satisfies
$\sup_y|\wt\rho(y)-\rho(y)|=o_p(1)$ and $\wh\E_p(\cdot\mid\x)$ satisfies
$\|\wh\E_p\{\a(Y)\mid\x\}-\Ep\{\a(Y)\mid\x\}\|_\infty=o_p(
n^{-1/4})$ for every bounded $\a(y)$ and $\x$, then
\bse
\sqrt{n}(\wh\bt-\bt)\to N\left[\0,\var\{\sqrt{\pi}\bphi\eff(R,RY,\X,\bt)\}\right]
\ese
in distribution as $n\to\infty$. Clearly, $\wh\bt$ achieves the semiparametric efficiency.
\end{theorem}

Theorem~\ref{th:general-eff} supports that $\var(\wh\bt)$ achieves the semiparametric efficiency bound for estimating $\bt$, as long as the estimator $\wh\bt$ is implemented with a consistent $\wt\rho$ and an $o_p(n^{-1/4})$-consistent $\wh\E_p$. This theoretical result differs from the results by \cite{lee2024doubly} on their singly flexible estimator $\wt\bt$. \cite{lee2024doubly} provided (i) the consistency of $\wt\bt$ given an arbitrary $\rho^*$ and an $o_p(n^{-1/4})$-consistent $\wh\E_p$, and (ii) the semiparametric efficiency of $\wt\bt$ given further that $\rho^*$ equals to the true $\rho$. We point out that (ii) is almost impossible in practice since there is almost no chance of an arbitrary chosen function $\rho^*$ being equal to the true function $\rho$, an object challenging to identify under our problem setting as discussed earlier. On the other hand, our proposed estimator $\wh\bt$ can achieve the semiparametric efficiency given an $o_p(n^{-1/4})$-consistent $\wh\E_p$ and a consistent $\wt\rho$, and we can easily provide $\wt\rho$ that is consistent for the true $\rho$ by implementing our proposed estimators $\wt\delta_0$ or $\wh\delta_0$.

In Theorems \ref{th:thetaeff} and \ref{th:general-eff}, we require the conditional expectation estimators $\wh\E_p(\cdot\mid\x)$ to attain the $o_p(n^{-1/4})$ convergence rate with respect to different norms in each theorem. Nonetheless, the requirements not significantly differ from each other. To see this, firstly, note that we regulate the gap between $\wh\E_p$ and $\E_p$ by the absolute value in Theorem \ref{th:thetaeff}, while in Theorem \ref{th:general-eff}, by the maximum of the sup-norms. We can treat these two norms interchangeably, as long as the parameter of interest has finite dimension which matches the dimension of the estimating equation, and the functions in the estimating equation are bounded uniformly on the support of variables. Secondly, we require the convergence rate is proportional to $\|a(y)\|_2$ in Theorem \ref{th:thetaeff}, while it is not in Theorem \ref{th:general-eff}. This gap is not significant as well. Specifically, consider a vector function $\b(y)$ not necessarily bounded and the convergence rate requirement in Theorem \ref{th:general-eff}. As long as we can identify the expression of $\|\b(y)\|_\infty$, letting $\a(y)=\b(y)/\|\b(y)\|_\infty$ transforms $\b(y)$ into the bounded function $\a(y)$. Now, suppose $\wh\E_p$ satisfies the convergence rate requirement in Theorem \ref{th:general-eff}, then plugging $\a(y)$ in the requirement leads us to $\|\wh\E_p\{\b(Y)\mid\x\}-\Ep\{\b(Y)\mid\x\}\|_\infty=o_p(n^{-1/4}\|\b(y)\|_\infty)$, which now is essentially similar to the one in Theorem \ref{th:thetaeff}.

\section{Discussion on Discrete $Y$}\label{sec:discreteY}

In Sections~\ref{sec:review} to \ref{sec:theory}, our proposed
methodology and theoretical
analysis have focused on scenarios where the outcome $Y$ is
continuous. We point out here that the established results
apply directly to the discrete setting.

Without loss of generality,  assume $Y$ takes values $1, \dots, K$.
To use our method for this case, all we need to do is to adopt
 bandwidths $h$ and $l$ that are less than 1, then no other changes
 are needed.
In fact, expressions will be simplified. For example,
to construct the efficient estimator of $\Pr(Y=y_0)$ in $\calQ$, first
calculate weights $\wh w_i = [\wh \E_p\{\wt
\rho^2(Y) + \pi / (1 - \pi) \wt \rho(Y) \mid \x_i\}]^{-1}$, and then we solve
\bse
\sumI \wh w_i\wh\E_p\{\wh a(Y)\wt\rho(Y)\mid\x_i\}\frac{r_i\I(y=y_i)}{\sumJ r_j\I(y=y_j)}=\I(y=y_0)
\ese
for $\wh a(y)$, then estimate $\Pr(Y=y_0)$ in $\calQ$ by
\bse
&&\frac1n\sumi\wt\rho(y_i)\left[\I(y_i=y_0)-\wh w_i\wh\E_p\{\wh a(Y)\wt\rho(Y)\mid\x_i\}\right]
+\frac1{N-n}\sum_{i=n+1}^N \wh w_i\wh\E_p\{\wh a(Y)\wt\rho(Y)\mid\x_i\}.
\ese

In fact, estimating $\rho(y)$ is equivalent to
estimating $\rho(k)$,
$k=1,\dots,K$. Thus, the proposed estimators $\wt\rho$ and $\wh\rho$
will achieve the parametric convergence rate $n^{-1/2}$.
Consequently, $\wh\rho(k)$'s,
will also retain the parametric convergence
rate. The
estimator $\wh\bt$, together with  $\wh\rho(k)$'s, will
preserve the established semiparametric efficiency, since
 the
 efficiency result of $\wh\bt$ provided in Theorem \ref{th:general-eff}
and of $\wh\rho(k)$'s in Corollary \ref{cor:effrho}
 is
guaranteed as long as $\wt\rho(y)$ is consistent.


\section{Simulation Studies}\label{sec:simu}

We present a simulation study to evaluate the performance of the
proposed estimators. The simulation is designed to first demonstrate
the density ratio estimators proposed in Section \ref{sec:rho}, then
compare various estimators of the mean and variance of $Y$ in
$\calQ$. We first introduce the estimators being compared, followed by
the process of generating the simulated data, and how the estimators
are implemented. The results are then illustrated using plots and
tables and discussed in detail.

We compare the following estimators of $\Eq\{ V(Y)\}$,
where $V(Y)=Y$ and $V(Y)=Y^2-\{\Eq(Y)\}^2$.
\begin{itemize}
    \item \texttt{Shift-dependent$^*$}: The shift dependent estimator $N^{-1}\sumI r_i\rho^*(y_i)V(y_i)/\pi$ naively incorporates a working model $\rho^*(y)$.
    \item \texttt{Doubly-flexible$^{*\star}$} and
      \texttt{Singly-flexible$^*$}: The estimators proposed by
      \cite{lee2024doubly} that incorporate a working model
      $\rho^*(y)$. \texttt{Doubly-flexible$^{*\star}$} further
      adopts a working regression model $\pyx^\star(y,\x,\bb)$, while
      \texttt{Singly-flexible$^*$} adopts a nonparametric regression
      estimator $\wh\E_p\{a(Y)\mid\x\}$.
    \item \texttt{Efficient}: Variations of the efficient estimators
      proposed in Section \ref{sec:general-eff} are implemented
      denoted as \texttt{Efficient\(^\sim\)}
      and \texttt{Efficient\(^\wedge\)},
      each corresponding to setting $\rho(y)$ to be $\wt\rho(y)$ and
      $\wh\rho(y)$. Note that these two estimators are asymptotically exactly the same.
    \item \texttt{Oracle}: The same efficient estimator is implemented
    with setting $\rho(y)$ to be the truth. Since the true $\rho(y)$ is unknown in practice,
      \texttt{Oracle} serves only as a benchmark.
\end{itemize}

The simulated data are generated with $N=500$ based on the following mechanism:
\[
R\sim {\rm Ber}(1/2),~ Y\mid R=1 \sim N(0, 2),~ Y\mid R=0 \sim N(\theta_1, \theta_2),~ \X\mid Y \sim N_3(\ba Y, \bI_3),
\]
where the mean of $Y$ in $\calQ$, $\theta_1=1$, the variance of $Y$ in $\calQ$, $\theta_2=1$, and \(\ba = (-0.5, 0.5, 1)\).

In our simulation study, the working model for $\rho(y)$ is specified
as $$\rho^*(y)=c^*\rho(y)\exp(0.2y + 0.1y^2),$$ where
$c^*=\pi/[N^{-1}\sumI\{r_i\rho(y_i)\exp(0.2y_i +
0.1y_i^2))\}]$.
This formulation incorrectly specifies $\rho(y)$. For the
  estimators $\wt\rho(y)$ and $\wh\rho(y)$ described in Section \ref{sec:rho}, we select the
bandwidth $h = 0.5n^{-1/16}$, where the order is chosen to satisfy the
regularity condition listed in Condition \ref{con:bandwidth}.
For the estimation of $\E\{b(\X)\mid y\}$ in the implementation of
\texttt{Efficient} and the estimators proposed by
\cite{lee2024doubly}, we adopt the Nadaraya-Watson estimator with
bandwidth $l = 1.5n^{-1/3}$. In addition, for the implementation of
\texttt{Efficient}, \texttt{Oracle}, and \texttt{Singly-flexible$^*$}, the
Nadaraya-Watson estimator of $\Ep\{a(Y)\mid\x\}$ is used as our
  machine-learning estimator,  with
bandwidth $3n^{-1/7}$.
For \texttt{Doubly-flexible$^{*\star}$}, the working model $\pyx^\star(y,\x,\bb)$ is set to be a normal regression model with the wrongly transformed covariate $$\left( X_1, \exp\left(\frac{X_2}{2}\right), \frac{X_3}{1+\exp(X_2)} + 10 \right).$$
The parameter $\bb$ is then estimated via maximum likelihood estimation.

\begin{figure}
    \centering
    \includegraphics[width=0.6\textwidth]{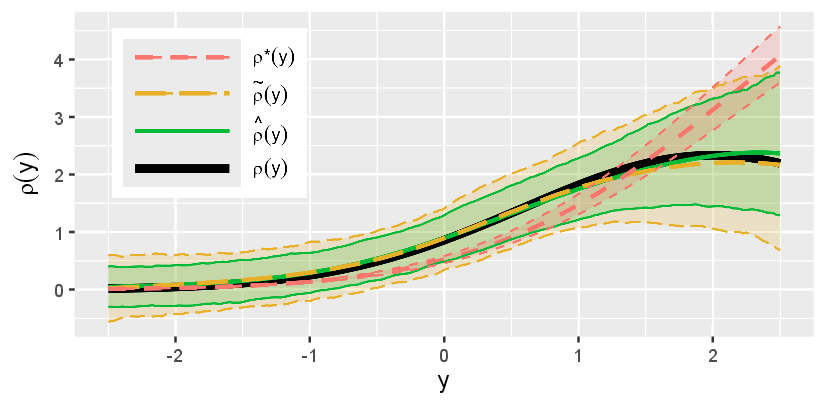}
    \caption{Plot of $\rho(y)$ estimates from 1000 simulation replicates.
    Lines: the average of 1000 estimates, Shades: the pointwise 90\% empirical confidence band of 1000 estimates.}
    \label{fig:rho}
\end{figure}

Figure \ref{fig:rho} presents the $\rho(y)$ estimates  from
1000 simulation replicates. The working model $\rho^*(y)$
significantly deviates from the true density ratio $\rho(y)$ as
suggested by its definition. In contrast, our proposed estimators
$\wt\rho(y)$  and $\wh\rho(y)$ proposed in Section \ref{sec:rho} provide a much more
accurate estimate of $\rho(y)$.
 We also observe
from the empirical confidence bands
that the variability of $\wh\rho(y)$ is much less compared to $\wt\rho(y)$,
i.e., $\wh\rho(y)$ provides  closer alignment with $\rho(y)$ than
$\wt\rho(y)$. This improvement
illustrates the effectiveness of our proposed self-improvement
procedure.

We now present the results for the estimation of the mean and variance of $Y$ in $\calQ$. The estimators are compared in terms of mean squared error multiplied by 100 (MSE), bias multiplied by 10 (Bias), standard error multiplied by 10 (SE), asymptotic relative efficiency calculated as the estimator's MSE divided by the \texttt{Oracle}'s MSE (ARE), and a coverage rate of asymptotic confidence interval with a confidence level $1-\alpha=0.95$ (CI) based on 1000 simulation replicates.

\begin{figure}
    \centering
    \includegraphics[width=0.6\textwidth]{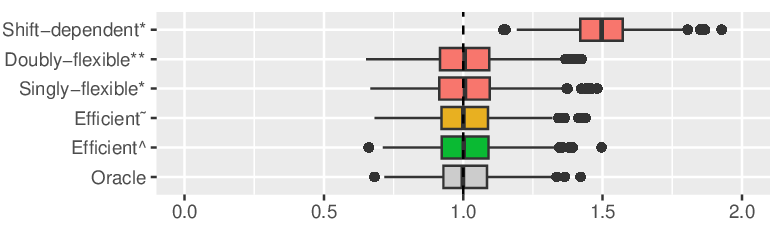}
    \caption{Boxplot of $\Eq(Y)$ estimates from 1000 simulation replicates.}
    \label{fig:mean}
\end{figure}

\begin{table}
    \caption{\small\label{tab:mean}Summary of $\Eq(Y)$ estimates from 1000 simulation replicates.}
    \centering
    \begin{tabular}{l|rrrrr}
        \hline
        Estimator & MSE & Bias & SE & ARE & CI\\
        \hline
        \texttt{Shift-dependent$^*$} & 25.8428 & 4.9437 & 1.1844 & 19.605 & 0.220 \\
        \texttt{Doubly-flexible$^{*\star}$} & 1.8189 & 0.0838 & 1.3461 & 1.380 & 0.951 \\
        \texttt{Singly-flexible$^*$} & 1.8039 & 0.0933 & 1.3399 & 1.368 & 0.952 \\
        \texttt{Efficient$^\sim$} & 1.4171 & 0.0620 & 1.1888 & 1.075 & 0.961 \\
        \texttt{Efficient$^\wedge$} & 1.4096 & 0.0649 & 1.1855 & 1.069 & 0.966 \\
        \texttt{Oracle} & 1.3182 & 0.0403 & 1.1474 & 1.000 & 0.961 \\
        \hline
    \end{tabular}
\end{table}

Figure \ref{fig:mean} and Table \ref{tab:mean} illustrate the
performance of estimates based on 1000 simulation
replicates. \texttt{Shift-dependent$^*$} exhibits significant bias,
and consequently, its CI falls far short of the nominal 95\%
  level. In contrast, both \texttt{Doubly-flexible$^{*\star}$} and
\texttt{Singly-flexible$^*$} by \cite{lee2024doubly} provide
relatively consistent estimates with improved Bias and
CI. \texttt{Efficient$^\sim$} and \texttt{Efficient$^\wedge$},
implemented using our proposed method, show an improvement in MSE
compared to the estimators by \cite{lee2024doubly}. Remarkably, the ARE's of
\texttt{Efficient$^\sim$} and \texttt{Efficient$^\wedge$} are close to 1, indicating that these estimators exhibit
efficiency similar to \texttt{Oracle} which uses the
true $\rho(y)$ which is unknown in practice. These
results provide empirical support for Theorem
\ref{th:general-eff}. Furthermore, the confidence intervals computed
using the asymptotic distribution provided in Theorem
\ref{th:general-eff} closely align with the nominal 95\% rate, further
validating the accuracy of our asymptotic results.

\begin{figure}
    \centering
    \includegraphics[width=0.6\textwidth]{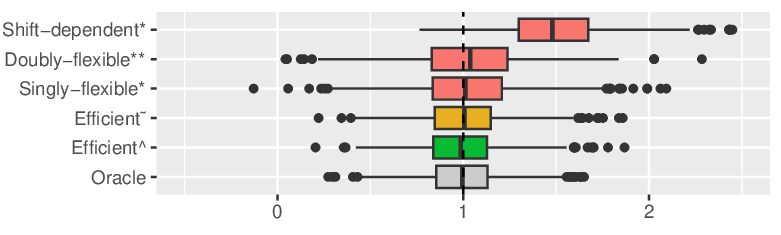}
    \caption{Boxplot of $\var_{\rm q}(Y)$ estimates from 1000 simulation replicates.}
    \label{fig:variance}
\end{figure}

\begin{table}
    \caption{\small\label{tab:variance}Summary of $\var_{\rm q}(Y)$ estimates from 1000 simulation replicates.}
    \centering
    \begin{tabular}{l|rrrrr}
        \hline
        Estimator & MSE & Bias & SE & ARE & CI \\
        \hline
        \texttt{Shift-dependent$^*$} & 31.9295 & 4.9199 & 2.7806 & 6.730 & 0.819 \\
        \texttt{Doubly-flexible$^{*\star}$} & 9.8121 & 0.3287 & 3.1167 & 2.068 & 0.959 \\
        \texttt{Singly-flexible$^*$} & 8.7310 & 0.2380 & 2.9467 & 1.840 & 0.957 \\
        \texttt{Efficient$^\sim$} & 5.3925 & 0.0281 & 2.3232 & 1.137 & 0.965 \\
        \texttt{Efficient$^\wedge$} & 5.1272 & -0.1510 & 2.2604 & 1.081 & 0.968 \\
        \texttt{Oracle} & 4.7441 & -0.0969 & 2.1770 & 1.000 & 0.961 \\
        \hline
    \end{tabular}
\end{table}

Figure \ref{fig:variance} and Table \ref{tab:variance} present the
boxplot and the summary table of the variance estimates from 1000
simulation replicates. Similar to the mean estimation results,
\texttt{Shift-dependent$^*$} estimator exhibits significant bias, with
its CI again failing to reach the nominal 95\% level. Also,
both \texttt{Doubly-flexible$^{*\star}$} and
\texttt{Singly-flexible$^*$} provide relatively consistent estimates
with lower Bias and improved CI. Notably, \texttt{Efficient$^\sim$}
and \texttt{Efficient$^\wedge$} demonstrate a more pronounced
improvement in MSE compared to the estimators by
\cite{lee2024doubly}. In fact, the ARE's of both
\texttt{Efficient$^\sim$} and \texttt{Efficient$^\wedge$} are close to 1,
indicating that the MSE's of these estimators are close that of \texttt{Oracle}. Moreover, the confidence intervals
based on the asymptotic distribution derived in Theorem
\ref{th:general-eff} once again demonstrate alignment with the nominal
95\% level, reinforcing the validity of our theoretical results.

\section{Real Data Applications}\label{sec:data}

\subsection{Plankton counting with computer vision}

Now we analyze a real-world dataset and demonstrate an improvement in
inference by our proposal from the existing methods. Our goal is to
count the number of planktons from the images taken by the Imaging
FlowCytobot, an automated submersible flow cytometry system at Woods
Hole Oceanographic Institution \citep{olson2003automated, orenstein2015whoi}.
The dataset consists of the images of matters
captured by the FlowCytobot from 2006 to 2014, the number of planktons
in 2014 was predicted based on a ResNet trained on the dataset from
2016 to 2013. The distribution of the numbers of planktons  shifted
from 2013 to 2014, mainly due to the change in the base frequencies of
plankton observations \citep{angelopoulos2023prediction}. Accordingly,
we consider the 2013 dataset with 421,238 labeled observations as
taken from $\calP$, and the 2014 dataset with 329,832 unlabeled
observations as taken from $\calQ$.

To estimate the proportion of the planktons and compare inference
results, we implemented the following methods: the prediction powered
inference by \cite{angelopoulos2023prediction} (\texttt{PPI$^*$}), the
shift dependent estimator $N^{-1}\sumI r_i\rho^*(y_i)y_i/\pi$ where
$y_i$ is coded $1$ if the image is a plankton or $0$ otherwise
(\texttt{Shift-dependent$^*$}), the singly flexible estimator by
\cite{lee2024doubly} (\texttt{Singly-flexible$^*$}), and our proposed
efficient estimators (\texttt{Efficient$^\sim$} and
\texttt{Efficient$^\wedge$}). To implement the existing methods, we
constructed a naive density ratio estimator using the true labels
$Y_i$ and the predicted labels $\wh Y_i$ in $\calP$, specifically,
$[\rho^*]_k=[\qy^*]_k/[\wh p_Y]_k$ where $\qy^*=\wh C^{-1}\wh q_X$, $[\wh C]_{k+1,l+1} =
\sum_{i=1}^n\I(\wh Y_i=k, Y_i=l)/\sum_{i=1}^n\I(Y_i=l)$, $[\wh
q_X]_{k+1}=(N-n)^{-1}\sum_{i=n+1}^N\I(\wh Y_i=k)$, and $[\wh
p_Y]_{k+1}=n^{-1}\sum_{i=1}^n\I(Y_i=k)$.
We constructed the conditional expectation estimators $\wh \E_p(\cdot\mid\x)$ using the last softmax layer of the ResNet, then used to implement the singly-flexible and our efficient estimators.

\begin{table}
    \caption{\small\label{tab:plankton}Summary of inference on the number of planktons.}
    \centering
    \begin{tabular}{l|rrrr}
        \hline
        Estimator & $\wh\theta$ & $\wh{\text{sd}}(\wh\theta)$ & CI Width & CI \\
        \hline
        \texttt{PPI$^*$}             & 39176.00    & 2159.734   & 8466.000  & [34943.00, 43409.00] \\
        \texttt{Shift-dependent$^*$} & 39152.00    & 147.521    & 578.271   & [38862.87, 39441.14] \\
        \texttt{Singly-flexible$^*$} & 39152.01    & 168.780    & 661.606   & [38821.21, 39482.81] \\
        \texttt{Efficient$^\sim$}    & 39152.01    & 110.477    & 433.062   & [38935.48, 39368.54] \\
        \texttt{Efficient$^\wedge$}  & 39152.01    & 110.477    & 433.062   & [38935.48, 39368.54] \\
        \hline
    \end{tabular}
\end{table}

The inference results are provided in Table \ref{tab:plankton}. We
report the point estimate $\wh\theta$, the estimated standard error
$\wh{\text{sd}}(\wh\theta)$, and the width of the 95\% confidence
interval, and its lower and upper bound. Firstly, the \texttt{PPI$^*$}
provided the widest confidence interval and its width was drastically
wider than the rest of the implemented methods. Other than the
\texttt{PPI$^*$}, all methods provided almost the same point
estimates. In terms of the width of CI, our proposed methods
\texttt{Efficient$^\sim$} and \texttt{Efficient$^\wedge$} provided
narrower confidence intervals compared to the
\texttt{Shift-dependent$^*$} and \texttt{Singly-flexible$^*$}. In this
analysis, the \texttt{Efficient$^\sim$} and
\texttt{Efficient$^\wedge$} resulted in the same interval estimate.

\subsection{Temperature forecast with humidity}

\begin{figure}
    \centering
    \includegraphics[width=0.6\textwidth]{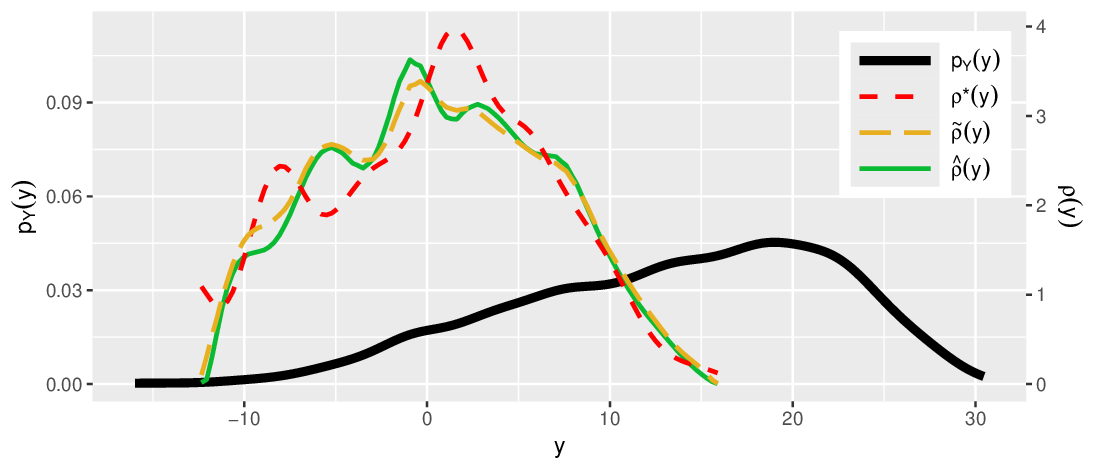}
    \caption{Temperature distribution in $\calP$ and the density ratio estimates.}
    \label{fig:temp}
\end{figure}

We now analyze the weather dataset of Szeged, Hungary from 2006 to
2016, and compare the performance of our proposed methods against the
existing methods. The dataset is publicly available at the following
link:
\url{https://www.kaggle.com/datasets/budincsevity/szeged-weather}. Several
weather characteristics of Szeged were recorded including temperature,
humidity, precipitation, and visibility. Our analysis goal is to
forecast the average temperature of Szeged in November and December
using humidity \citep{kim2024retasa}. In this dataset, we consider the
daily average humidity as a covariate and the daily average
temperature as a response, and take 3,348 observations between January
and October as taken from $\calP$ and 671 observations between
November and December as taken from $\calQ$. The
  empirical distribution of the average temperature in $\calP$ is
  visualized as the black curve in Figure \ref{fig:temp} based on
  its kernel density estimate.

To forecast the average temperature of Szeged in November and December, we implemented the following methods: the shift dependent estimator $N^{-1}\sumI r_i\rho^*(y_i)y_i/\pi$, the doubly-flexible (\texttt{Doubly-flexible$^{*\star}$}) and singly-flexible (\texttt{Singly-flexible$^*$}) estimators by \cite{lee2024doubly}, and our proposed efficient estimators (\texttt{Efficient$^\sim$} and \texttt{Efficient$^\wedge$}). To implement the existing methods, we created an arbitrary function $\rho^*(y)$ which is visualized in Figure \ref{fig:temp}. We adopted a normal regression model as a working model for $\pyx$, and used the Nadaraya-Watson estimator based on the observations in $\calP$ as $\wh\E_p(\cdot\mid\x)$. Further, the bandwidth $h$ is chosen as $1.5n^{-1/16}$ and $l$ as $3n^{-1/3}$.

\begin{table}
\caption{\small\label{tab:temp}Summary of inference on the average temperature.}
\centering
\begin{tabular}{l|rrrr}
    \hline
    Estimator & $\wh\theta$ & $\wh{\text{sd}}(\wh\theta)$ & CI Width & CI \\
    \hline
    \texttt{Shift-dependent$^*$}        & 5.538  & 0.153  & 0.602  & [5.238, 5.839] \\
    \texttt{Doubly-flexible$^{*\star}$} & 4.267  & 0.504  & 1.976  & [3.280, 5.255] \\
    \texttt{Singly-flexible$^*$}        & 3.998  & 0.160  & 0.629  & [3.683, 4.312] \\
    \texttt{Efficient$^\sim$}           & 4.086  & 0.123  & 0.482  & [3.845, 4.327] \\
    \texttt{Efficient$^\wedge$}         & 4.131  & 0.126  & 0.493  & [3.884, 4.377] \\
    \hline
\end{tabular}
\end{table}

Figure \ref{fig:temp} visualizes the density ratio
estimates $\wt\rho$ and $\wh\rho$, and Table \ref{tab:temp} summarizes
the inference results. We again report the point estimate $\wh\theta$,
the estimated standard error $\wh{\text{sd}}(\wh\theta)$, and the
width of the 95\% confidence interval, and its lower and upper
bound. The point estimate from \texttt{Shift-dependent$^*$}, 5.538,
was much bigger than the rest of the methods, which were around
4.1. Also, we do not see much overlap of the CI of
\texttt{Shift-dependent$^*$} with the rest. On the other hand, we
notice that the CI width of \texttt{Doubly-flexible$^{*\star}$} was
much wider than \texttt{Singly-flexible$^*$} and our efficient
estimators, possibly due to the departure of the chosen working model
from the underlying true distribution. \texttt{Singly-flexible$^*$}
narrows the CI width from \texttt{Doubly-flexible$^{*\star}$} by
adopting an estimator of $\Ep(\cdot\mid\x)$ that is more flexible
hence more likely to cover the truth, but our efficient estimators
\texttt{Efficient$^\sim$} and \texttt{Efficient$^\wedge$} improve even
more and provide the narrowest confidence intervals.

\section{Conclusion}\label{sec:disc}

In this paper, we consider distribution shift, specifically label shift, and we propose an efficient inference procedure.
The presence of distribution shift introduces additional complexity compared to cases without such shifts.
Specifically, rather than simply predicting the conditional mean
function as $\wh Y$, one must predict $\wh Y$ as $\wh \b(\X)$, still a
function of the feature $\X$ but with a significantly more intricate
structure. 
A crucial element for achieving efficient inference is modeling the
outcome density ratio between labeled and unlabeled data $\rho(y)$. 
To this end, we introduce a progressive estimation process comprising
three stages: an initial heuristic guess, a consistent estimation, and
finally, an efficient estimation. 
This self-driven, iterative approach is unconventional in the
statistical literature and may hold independent interest beyond its
application in this context. 

\section*{Supplementary Material}

The Supplementary Material includes all of the technical details.

\section*{Conflict of Interest}

The authors report there are no competing interests to declare.

\begin{acks}[Acknowledgments]
All correspondences should be addressed to Jiwei Zhao (the corresponding author) at \href{Email:jiwei.zhao@wisc.edu}{jiwei.zhao@wisc.edu}.
\end{acks}

\begin{funding}
The research is supported in part by NSF (DMS 1953526, 2122074, 2310942), NIH (R01DC021431) and the American Family Funding Initiative of UW-Madison.
\end{funding}


\bibliographystyle{imsart-nameyear}
\bibliography{labelshiftpred}


\clearpage
\begin{supplement}
\stitle{}
\setcounter{page}{1}
\setcounter{equation}{0}\renewcommand{\theequation}{S.\arabic{equation}}
\setcounter{section}{0}\renewcommand{\thesection}{S.\arabic{section}}
\setcounter{subsection}{0}\renewcommand{\thesubsection}{S.\arabic{subsection}}

\section{Derivation of the EIF for $\delta_0$}\label{supp:sec:deltaEIF}

The efficient influence function for $\bt$ that satisfies $\Eq\{\U(Y,\X,\bt)\}=0$ is, by Proposition S.3 in the supplement of \cite{lee2024doubly},
\bse
\phi\eff(r,ry,\x)
&=&\frac{r}{\pi}\rho(y)\A\left[\U(y,\x,\bt)-\frac{\Ep\{\U(Y,\x,\bt)\rho^2(Y)+\a(Y)\rho(Y)\mid\x\}}{\Ep\{\rho^2(Y)+\pi/(1-\pi)\rho(Y)\mid\x\}}\right]\\
&&+\frac{1-r}{1-\pi}\frac{\A\Ep\{\U(Y,\x,\bt)\rho^2(Y)+\a(Y)\rho(Y)\mid\x\}}{\Ep\{\rho^2(Y)+\pi/(1-\pi)\rho(Y)\mid\x\}},
\ese
where $\A=[\Eq\{\partial \U(Y,\X,\bt)/\partial\bt\trans\}]^{-1}$ and $\a(y)$ satisfies
\bse
\E\left[\frac{\Ep\{\U(Y,\X,\bt)\rho^2(Y)+\a(Y)\rho(Y)\mid\X\}}{\Ep\{\rho^2(Y)+\pi/(1-\pi)\rho(Y)\mid\X\}}\mid y\right]
=\E\{\U(y,\X,\bt)\mid y\}.
\ese
Note that $\delta_0=\Eq\{K_h(Y-y_0)\}$ satisfies $\Eq\{U(Y,\X,\delta_0)\}=0$ with $U(y,\x,\delta)=K_h(y-y_0)-\delta$. Then $\phi\eff$ reduces to
\bse
\phi\eff(r,ry,\x)
&=&-\frac{r}{\pi}\rho(y)\left[K_h(y-y_0)-\delta_0-\frac{\Ep\{K_h(Y-y_0)\rho^2(Y)-\delta_0\rho^2(Y)+a(Y)\rho(Y)\mid\x\}}{\Ep\{\rho^2(Y)+\pi/(1-\pi)\rho(Y)\mid\x\}}\right]\\
&&-\frac{1-r}{1-\pi}\frac{\Ep\{K_h(Y-y_0)\rho^2(Y)-\delta_0\rho^2(Y)+a(Y)\rho(Y)\mid\x\}}{\Ep\{\rho^2(Y)+\pi/(1-\pi)\rho(Y)\mid\x\}},
\ese
where $a(y)$ satisfies
\bse
\E\left[\frac{\Ep\{K_h(Y-y_0)\rho^2(Y)-\delta_0\rho^2(Y)+a(Y)\rho(Y)\mid\X\}}{\Ep\{\rho^2(Y)+\pi/(1-\pi)\rho(Y)\mid\X\}}\mid y\right]
=K_h(y-y_0)-\delta_0.
\ese
Now, let $\wt a(y)=a(y)+K_h(y-y_0)\rho(y)+\delta_0\pi/(1-\pi)\rho(y)$, then $\phi\eff$ further reduces to
\bse
\phi\eff(r,ry,\x)
&=&-\frac{r}{\pi}\rho(y)\left[K_h(y-y_0)-\frac{\Ep\{\wt a(Y)\rho(Y)\mid\x\}}{\Ep\{\rho^2(Y)+\pi/(1-\pi)\rho(Y)\mid\x\}}\right]\\
&&-\frac{1-r}{1-\pi}\left[\frac{\Ep\{\wt a(Y)\rho(Y)\mid\x\}}{\Ep\{\rho^2(Y)+\pi/(1-\pi)\rho(Y)\mid\x\}}-\delta_0\right],
\ese
where $\wt a(y)$ satisfies
\bse
\E\left[\frac{\Ep\{\wt a(Y)\rho(Y)\mid\X\}}{\Ep\{\rho^2(Y)+\pi/(1-\pi)\rho(Y)\mid\X\}}\mid y\right]=K_h(y-y_0).
\ese
\hfill$\blacksquare$


\section{Proof of Lemma \ref{lem:i}}\label{supp:sec:prooflemma1}
We first show that $\calI^{*\dag}$ is invertible at $v_h$ by contradiction. Suppose there are $a_1$ and $a_2$ such that $a_1\neq a_2$ and $\calI^{*\dag}(a_1)=\calI^{*\dag}(a_2)=v_h$. Then by Conditions \ref{con:rho} and \ref{con:pyx}, we have
\bse
\Ep^\dag\{a_1(Y)\rho^*(Y)\mid\x\}\neq\Ep^\dag\{a_2(Y)\rho^*(Y)\mid\x\}.
\ese
On the other hand, the efficient score under the posited models is
\bse
\phi\eff^{*\dag}(r,ry,\x)
&=&-\frac{r}{\pi}\rho^*(y)\left[K_h(y-y_0)
-\frac{\Ep^\dag\{a(Y)\rho^*(Y)\mid\x\}}
{\Ep^\dag\{\rho^{*2}(Y)+\pi/(1-\pi)\rho^*(Y)\mid\x\}}\right]\\
&&-\frac{1-r}{1-\pi}
\left[\frac{\Ep^\dag\{a(Y)\rho^*(Y)\mid\x\}}
{\Ep^\dag\{\rho^{*2}(Y)+\pi/(1-\pi)\rho^*(Y)\mid\x\}}-\delta_0\right],
\ese
where $a(y)$ satisfies $\calI^{*\dag}(a)=v_h$. Then letting $a=a_1$
and $a=a_2$ gives two distinct efficient scores, which contradicts the
uniqueness of the efficient score. Therefore, $a_h^{*\dag}$ is a
unique solution to $\calI^{*\dag}(a)=v_h$, hence $\calI^{*\dag}$ is
invertible at $v_h$.

Now, we show $\calI^{*\dag}$ is invertible at any $v$ in the range of
$\calI^{*\dag}$ by contradiction. Suppose there are $a_1$ and $a_2$
such that $a_1\neq a_2$ and
$\calI^{*\dag}(a_1)=\calI^{*\dag}(a_2)=v$. This yields
$\calI^{*\dag}(a_h^{*\dag}+a_1-a_2)=\calI^{*\dag}(a_h^{*\dag})+\calI^{*\dag}(a_1)-\calI^{*\dag}(a_2)=v_h$. Then
letting $a=a_h^{*\dag}$ and $a=a_h^{*\dag}+a_1-a_2$ gives two distinct
efficient scores, and this again contradicts the uniqueness of the
efficient score. Hence, there is a unique solution to
$\calI^{*\dag}(a)=v$. Therefore, $\calI^{*\dag}$ is invertible.

Next, we show the existence of the constants $C_1,C_2$. Since
\bse
\|\calI^{*\dag}(a)\|_2^2=\int\left\{\int a(t)u^{*\dag}(t,y)dt\right\}^2dy\leq\|a\|_2^2\int\|u^{*\dag}(\cdot,y)\|_2^2dy
\ese
there exists a constant $C_1$ such that $0<C_1<\infty$ and $\|\calI^{*\dag}(a)\|_2\leq C_1\|a\|_2$ by Conditions \ref{con:u} and \ref{con:density}. In addition, we have $\|\calI^{*\dag-1}(v)\|_2\leq C_2\|v\|_2$ for some constant $C_2$ such that $0<C_2<\infty$ by the bounded inverse theorem, since $\calI^{*\dag}$ is bounded and invertible.
\hfill$\blacksquare$


\section{Proof of Theorem \ref{th:theta}}\label{supp:sec:proofthm1}
Let us define
\bse
b^*(\x,a,\Ep)
&\equiv&\frac{\Ep\{a(Y)\rho^*(Y)\mid\x\}}{\Ep\{\rho^{*2}(Y)+\pi/(1-\pi)\rho^*(Y)\mid\x\}},\\
\calI_{l,i}^{*\dag}(a)(y)
&\equiv&\frac{r_i}{\pi}\wt K_l(y-y_i)b^*(\x_i,a,\Ep^\dag),\\
\wh\calI^{*\dag}(a)(y)
&\equiv&N^{-1}\sumI\frac{r_i}{\pi}\wt K_l(y-y_i)b^*(\x_i,a,\Ep^\dag),\\
v_{h,l,i}(y)
&\equiv&\frac{r_i}{\pi}\wt K_l(y-y_i)K_h(y-y_0),\\
\wh v(y)
&\equiv&N^{-1}\sumI v_{h,l,i}(y),\\
d_{h,l,i}(y)
&\equiv&\calI^{*\dag-1}\{v_{h,l,i}-\calI_{l,i}^{*\dag}(a_h^{*\dag})\}(y).
\ese
Note that
\be\label{eq:eb}
\E\{b^*(\X,a_h^{*\dag},\Ep^\dag)\mid y\}
=\frac{\calI^{*\dag}(a_h^{*\dag})(y)}{\py(y)}
=\frac{v_h(y)}{\py(y)}
=K_h(y-y_0).
\ee

First, we derive the norms of some functions. Since $\calI^{*\dag-1}$
is linear and bounded by Lemma \ref{lem:i} and
$\|v_h\|_2=O(h^{-1/2})$,
\be\label{eq:a}
\|a_h^{*\dag}\|_2=\|\calI^{*\dag-1}(v_h)\|_2=O(h^{-1/2})
\ee
by Lemma \ref{lem:i}. In addition,
with $V_{h,l,i}(y)\equiv\frac{R_i}{\pi}\wt K_l(y-Y_i)K_h(y-y_0)$,
\bse
\|\E\{\calI^{*\dag-1}(V_{h,l,i})\}-\calI^{*\dag-1}(v_h)\|_2
&=&\|\calI^{*\dag-1}\{\E(V_{h,l,i})\}-\calI^{*\dag-1}(v_h)\|_2\\
&=&\|\calI^{*\dag-1}\{\E(V_{h,l,i})-v_h\}\|_2\\
&=&O(h^{-1/2}l^m),
\ese
where the last equality holds by Lemma \ref{lem:i} and
$\|\E(V_{h,l,i})-v_h\|_2=O(h^{-1/2}l^m)$ under Conditions
\ref{con:density} and \ref{con:kernel}. Also,
\bse
&&\|\E\{(\calI^{*\dag-1}\circ\calI_{l,i}^{*\dag})(a_h^{*\dag})\}-a_h^{*\dag}\|_2\\
&=&\|\E\{(\calI^{*\dag-1}\circ\calI_{l,i}^{*\dag})(a_h^{*\dag})\}-\calI^{*\dag-1}(v_h)\|_2\\
&=&\|\calI^{*\dag-1}[\E\{\calI_{l,i}^{*\dag}(a_h^{*\dag})\}-v_h]\|_2\\
&=&O\left[\left\|\E\left\{\frac{R}{\pi}\wt K_l(y-Y)b^*(\X,a_h^{*\dag},\Ep^\dag)\right\}-v_h\right\|_2\right]\\
&=&O\left[\|\py(y)\E\{b^*(\X,a_h^{*\dag},\Ep^\dag)\mid y\}+O(h^{-1/2}l^m)-v_h\|_2\right]\\
&=&O(h^{-1/2}l^m),
\ese
where the third equality holds by Lemma \ref{lem:i}, the fourth
equality holds by \eqref{eq:a}, Conditions \ref{con:density}, and
\ref{con:kernel}, and the last equality holds by \eqref{eq:eb}. Then,
\be\label{eq:edelta}
\|\E(D_{h,l,i})\|_2
&=&\|\E\{\calI^{*\dag-1}(V_{h,l,i})\}-\E\{(\calI^{*\dag-1}\circ\calI_{l,i}^{*\dag})(a_h^{*\dag})\}-\{\calI^{*\dag-1}(v_h)-a_h^{*\dag}\}\|_2\n\\
&\leq&\|\E\{\calI^{*\dag-1}(V_{h,l,i})\}-\calI^{*\dag-1}(v_h)\|_2+\|\E\{(\calI^{*\dag-1}\circ\calI_{l,i})(a_h^{*\dag})\}-a_h^{*\dag}\|_2\n\\
&=&O(h^{-1/2}l^m).
\ee
In addition,
\be\label{eq:deltaclt}
&&\left\|N^{-1}\sumI d_{h,l,i}-\E(D_{h,l,i})\right\|_2\n\\
&=&\left\|N^{-1}\sumI\calI^{*\dag-1}\{v_{h,l,i}-\calI_{l,i}^{*\dag}(a_h^{*\dag})\}-\E\left[\calI^{*\dag-1}\{V_{h,l,i}-\calI_{l,i}^{*\dag}(a_h^{*\dag})\}\right]\right\|_2\n\\
&=&\left\|\calI^{*\dag-1}\left[N^{-1}\sumI v_{h,l,i}-\E(V_{h,l,i})-N^{-1}\sumI\calI_{l,i}^{*\dag}(a_h^{*\dag})+\E\{\calI_{l,i}^{*\dag}(a_h^{*\dag})\}\right]\right\|_2\n\\
&=&O\left[\left\|N^{-1}\sumI v_{h,l,i}-\E(V_{h,l,i})\right\|_2+\left\|N^{-1}\sumI\calI_{l,i}^{*\dag}(a_h^{*\dag})-\E\{\calI_{l,i}^{*\dag}(a_h^{*\dag})\}\right\|_2\right]\n\\
&=&O_p\{h^{-1/2}(nl)^{-1/2}\},
\ee
where the second equality holds since $\calI^{*\dag-1}$ is linear by
Lemma \ref{lem:i}, the third equality holds since $\calI^{*\dag-1}$ is
bounded by Lemma \ref{lem:i}, and the last equality holds by
\eqref{eq:a} and Condition \ref{con:kernel}.
On the other hand, for any function $a(y)$,
\be\label{eq:ihat}
\|(\wh\calI^{*\dag}-\calI^{*\dag})(a)\|_2
&=&\left\|N^{-1}\sumI\frac{r_i}{\pi}\wt K_l(y-y_i)b^*(\x_i,a,\Ep^\dag)-\py(y)\E\{b^*(\x,a,\Ep^\dag)\mid y\}\right\|_2\n\\
&=& 
O_p\left[\{(nl)^{-1/2}+l^m\}\|a\|_2\right]=o_p(n^{-1/4}\|a\|_2)
\ee
under Conditions \ref{con:u}-\ref{con:bandwidth}.
Similarly,
\be\label{eq:vhat}
\|\wh v-v_h\|_2= O_p\left\{(nl)^{-1/2}+l^m\right\}O(h^{-1/2})=o_p(n^{-1/4}h^{-1/2}).
\ee

Now, we derive the asymptotic form of $b^*(\x,\wh a^{*\dag},\Ep^\dag)$. $\wh a^{*\dag}$ can be written as, since $\calI^{*\dag-1}$ is linear and bounded by Lemma \ref{lem:i},
\be\label{eq:ahat}
\wh a^{*\dag}(y)
&=&\wh\calI^{*\dag-1}(\wh v)(y)\n\\
&=&\{\calI^{*\dag}+(\wh\calI^{*\dag}-\calI^{*\dag})\}^{-1}(\wh v)(y)\n\\
&=&\{\calI^{*\dag-1}-\calI^{*\dag-1}\circ(\wh\calI^{*\dag}-\calI^{*\dag})\circ\calI^{*\dag-1}\}\{v_h+(\wh v-v_h)\}(y)+o_p(n^{-1/2}h^{-1/2})\n\\
&=&a_h^{*\dag}(y)+\calI^{*\dag-1}(\wh v-v_h)(y)-\{\calI^{*\dag-1}\circ(\wh\calI^{*\dag}-\calI^{*\dag})\}(a_h^{*\dag})(y)+o_p(n^{-1/2}h^{-1/2})\n\\
&=&a_h^{*\dag}(y)+\calI^{*\dag-1}\{\wh v-\wh\calI^{*\dag}(a_h^{*\dag})\}(y)+o_p(n^{-1/2}h^{-1/2})\n\\
&=&a_h^{*\dag}(y)+N^{-1}\sumI d_{h,l,i}(y)+o_p(n^{-1/2}h^{-1/2})
\ee
uniformly in $y$ by Condition \ref{con:density}. Here, the third
equality holds by \eqref{eq:ihat} and
\bse
\|\wh v\|_2\leq\|v_h\|_2+\|\wh v-v_h\|_2=O(h^{-1/2})+o_p(n^{-1/4}h^{-1/2})=O_p(h^{-1/2})
\ese
by \eqref{eq:vhat}, and the fourth equality holds by
$a_h^{*\dag}(y)=\calI^{*\dag-1}(v_h)(y)$, \eqref{eq:ihat}, and
\eqref{eq:vhat}.
Then,
\be\label{eq:ahat-a}
\|\wh a^{*\dag}-a_h^{*\dag}\|_2=O(h^{-1/2}l^m)+O_p\{h^{-1/2}(nl)^{-1/2}\}+o_p(n^{-1/2}h^{-1/2})=o_p(n^{-1/4}h^{-1/2})
\ee
by \eqref{eq:edelta}, \eqref{eq:deltaclt},
\eqref{eq:ahat}, and Condition \ref{con:bandwidth},
and
\be\label{eq:ahat2}
\|\wh a^{*\dag}\|_2\leq\|a_h^{*\dag}\|_2+\|\wh a^{*\dag}-a_h^{*\dag}\|_2=O_p(h^{-1/2})
\ee
by \eqref{eq:a}.
Hence,
\be\label{eq:bhat}
&&b^*(\x,\wh a^{*\dag},\Ep^\dag)\\
&=&b^*(\x,a_h^{*\dag},\Ep^\dag)+b^*(\x,\wh a^{*\dag}-a_h^{*\dag},\Ep^\dag)\n\\
&=&b^*(\x,a_h^{*\dag},\Ep^\dag)+N^{-1}\sumI b^*(\x,d_{h,l,i},\Ep^\dag)
+o_p(n^{-1/2}h^{-1/2})\n
\ee
uniformly in $\x$ by Condition \ref{con:density}, where 
the last equality holds by \eqref{eq:ahat}.

Now, we analyze $\wt\delta_0$. From the definition of $\wt\delta_0$,
\be\label{eq:theta}
\wt\delta_0-\delta_0
&=&N^{-1}\sumI\left[\frac{r_i}{\pi}\rho^*(y_i)\{K_h(y_i-y_0)-b^*(\x_i,\wh a^{*\dag},\Ep^\dag)\}
+\frac{1-r_i}{1-\pi}\{b^*(\x_i,\wh a^{*\dag},\Ep^\dag)-\delta_0\}\right]\n\\
&=&-N^{-1}\sumI\phi\eff^{*\dag}(r_i,r_iy_i,\x_i)\n\\
&&-N^{-1}\sumI\left\{\frac{r_i}{\pi}\rho^*(y_i)-\frac{1-r_i}{1-\pi}\right\}
\{b^*(\x_i,\wh a^{*\dag},\Ep^\dag)-b^*(\x_i,a_h^{*\dag},\Ep^\dag)\}\n\\
&=&-N^{-1}\sumI\phi\eff^{*\dag}(r_i,r_iy_i,\x_i)-T+o_p(n^{-1/2}h^{-1/2}),
\ee
where
\bse
T&\equiv&N^{-2}\sumI\sumJ\left\{\frac{r_i}{\pi}\rho^*(y_i)-\frac{1-r_i}{1-\pi}\right\}b^*(\x_i,d_{h,l,j},\Ep^\dag).
\ese
By the property of the U-statistic, $T$ can be written as
\be\label{eq:t1}
&&T\\
&=&N^{-1}\sumI
\left\{\frac{r_i}{\pi}\rho^*(y_i)-\frac{1-r_i}{1-\pi}\right\}
\E\{b^*(\x_i,D_{h,l,j},\Ep^\dag)\mid r_i,r_iy_i,\x_i\}\n\\
&&+N^{-1}\sumJ
\E\left[\left\{\frac{R_i}{\pi}\rho^*(Y_i)-\frac{1-R_i}{1-\pi}\right\}
b^*(\X_i,d_{h,l,j},\Ep^\dag)\mid r_j,r_jy_j,\x_j\right]\n\\
&&-N^{-1/2}\E\left[\left\{\frac{R_i}{\pi}\rho^*(Y_i)-\frac{1-R_i}{1-\pi}\right\}
b^*(\X_i,D_{h,l,j},\Ep^\dag)\right]+O_p(N^{-1}h^{-1/2})\n\\
&=&N^{-1}\sumJ\E\left[\left\{\frac{R_i}{\pi}\rho^*(Y_i)-\frac{1-R_i}{1-\pi}\right\}
b^*(\X_i,d_{h,l,j},\Ep^\dag)\mid r_j,r_jy_j,\x_j\right]\n\\
&&+O_p(h^{-1/2}l^m+N^{-1}h^{-1/2})\n\\
&=&N^{-1}\sumJ\int\left[\rho^*(y)\E\{b^*(\X,d_{h,l,j},\Ep^\dag)\mid y\}\py(y)
-\E\{b^*(\X,d_{h,l,j},\Ep^\dag)\mid y\}\qy(y)\right]dy\n\\
&&+o_p(n^{-1/2}h^{-1/2})\n\\
&=&N^{-1}\sumJ\int\left\{\rho^*(y)-\rho(y)\right\}\calI^{*\dag}(d_{h,l,j})(y)dy+o_p(n^{-1/2}h^{-1/2})\n\\
&=&N^{-1}\sumJ\int\left\{\rho^*(y)-\rho(y)\right\}\{v_{h,l,j}(y)-\calI_{l,j}^{*\dag}(a_h^{*\dag})(y)\}dy+o_p(n^{-1/2}h^{-1/2})\n\\
&=&N^{-1}\sumJ\frac{r_j}{\pi}\int \wt K_l(y-y_j)\left\{\rho^*(y)-\rho(y)\right\}\{K_h(y-y_0)-b^*(\x_j,a_h^{*\dag},\Ep^\dag)\}dy+o_p(n^{-1/2}h^{-1/2}),\n
\ee
where the second equality holds because \eqref{eq:edelta} leads to
\bse
\E\{b^*(\x_i,D_{h,l,j},\Ep^\dag)\mid r_i,r_iy_i,\x_i\}=O(h^{-1/2}l^m),
\ese
and the third equality holds by Condition \ref{con:bandwidth}. Further,
\be\label{eq:t12}
&&\int \wt K_l(y-y_j)\left\{\rho^*(y)-\rho(y)\right\}\{K_h(y-y_0)-b^*(\x_j,a_h^{*\dag},\Ep^\dag)\}dy\n\\
&=&\int \wt K(t)\{\rho^*(y_j+lt)-\rho(y_j+lt)\}\{K_h(y_j+lt-y_0)-b^*(\x_j,a_h^{*\dag},\Ep^\dag)\}dt\n\\
&=&\{\rho^*(y_j)-\rho(y_j)\}\{K_h(y_j-y_0)-b^*(\x_j,a_h^{*\dag},\Ep^\dag)\}+
O(h^{-m-1}l^m)\n\\
&=&\{\rho^*(y_j)-\rho(y_j)\}\{K_h(y_j-y_0)-b^*(\x_j,a_h^{*\dag},\Ep^\dag)\}+o(n^{-1/2}h^{-1/2})
\ee
under Conditions \ref{con:rho}, \ref{con:kernel}, and
\ref{con:bandwidth}.
Therefore, \eqref{eq:theta}, \eqref{eq:t1}, and \eqref{eq:t12} 
lead to
\be\label{eq:wttheta}
&&\wt\delta_0-\delta_0\n\\
&=&-N^{-1}\sumI\phi\eff^{*\dag}(r_i,r_iy_i,\x_i)\n\\
&&-N^{-1}\sumI\frac{r_i}{\pi}\{\rho^*(y_i)-\rho(y_i)\}\{K_h(y_i-y_0)-b^*(\x_i,a_h^{*\dag},\Ep^\dag)\}+o_p(n^{-1/2}h^{-1/2})\n\\
&=&N^{-1}\sumI\frac{r_i}{\pi}\rho(y_i)\left\{K_h(y_i-y_0)-b^*(\x_i,a_h^{*\dag},\Ep^\dag)\right\}\n\\
&&+N^{-1}\sumI\frac{1-r_i}{1-\pi}\left\{b^*(\x_i,a_h^{*\dag},\Ep^\dag)-\delta_0\right\}+o_p(n^{-1/2}h^{-1/2}).
\ee
Note that
\bse
\var(\wt\delta_0-\delta_0)
&=&N^{-1}\E\left(\left[\frac{R_i}{\pi}\rho(Y_i)\left\{K_h(Y_i-y_0)-b^*(\X_i,a_h^{*\dag},\Ep^\dag)\right\}\right]^2\right)\\
&&+N^{-1}\E\left(\left[\frac{1-R_i}{1-\pi}\left\{b^*(\X_i,a_h^{*\dag},\Ep^\dag)-\delta_0\right\}\right]^2\right)\\
&&+ o[\{\min(n,N-n)\}^{-1/2}n^{-1/2}h^{-1}]\\
&=&n^{-1}\E\left(\frac{R_i}{\pi}\left[\rho(Y_i)\left\{K_h(Y_i-y_0)-b^*(\X_i,a_h^{*\dag},\Ep^\dag)\right\}\right]^2\right)\\
&&+(N-n)^{-1}\E\left[\frac{1-R_i}{1-\pi}\left\{b^*(\X_i,a_h^{*\dag},\Ep^\dag)-\delta_0\right\}^2\right]\\
&&+ o[\{\min(n,N-n)\}^{-1/2}n^{-1/2}h^{-1}]\\
&=&n^{-1}\Ep\left(\left[\rho(Y_i)\left\{K_h(Y_i-y_0)-b^*(\X_i,a_h^{*\dag},\Ep^\dag)\right\}\right]^2\right)\\
&&+(N-n)^{-1}\Eq\left[\left\{b^*(\X_i,a_h^{*\dag},\Ep^\dag)-\delta_0\right\}^2\right]\\
&&+ o[\{\min(n,N-n)\}^{-1/2}n^{-1/2}h^{-1}].
\ese
  In addition,
\bse
\delta_0
&=&\int K_h(y-y_0)\qy(y)dy\\
&=&\int K(t)\qy(y_0+ht)dt\\
&=&\qy(y_0)+\qy^{(m)}(y_0)\frac{\int t^mK(t)dt}{m!}h^m+O(h^{m+1})
\ese
under Conditions \ref{con:density}, \ref{con:kernel}, and \ref{con:bandwidth}.
\hfill$\blacksquare$


\section{Proof of Theorem \ref{th:rhomax}}\label{supp:sec:proofthm2}
Let
\bse
z_i\equiv\frac{r_i}{\pi}\rho(y_i)\left\{K_h(y_i-y_0)-b^*(\x_i,a_h^{*\dag},\Ep^\dag)\right\}
+\frac{1-r_i}{1-\pi}\left\{b^*(\x_i,a_h^{*\dag},\Ep^\dag)-\delta_0\right\}.
\ese
We have
\be\label{eq:bernstein1}
|z_i|
&=&O\left\{\frac{N}{n}(h^{-1}+h^{-1/2})+\frac{N}{N-n}(h^{-1/2}+1)\right\}\n\\
&\leq&C_1\left\{\frac{N}{n}h^{-1}+\frac{N}{\min(n,N-n)}h^{-1/2}\right\}
\ee
for some constant $C_1>0$ uniformly for $i$,
because $K_h(y_i-y_0)=O(h^{-1})$ by Condition \ref{con:kernel},
$b^*(\x_i,a_h^{*\dag},\Ep^\dag)=O(h^{-1/2})$ since $\rho^*$ is
bounded above and bounded away from zero by Condition \ref{con:rho}
and $\|a_h^{*\dag}\|_2=O(h^{-1/2})$ by \eqref{eq:a},
$\delta_0=\Eq\{K_h(Y-y_0)\}=O(1)$, and $h=o(1)$ by Condition
\ref{con:bandwidth}. In addition,
\be\label{eq:bernstein2}
&&\var(Z_i)\n\\
&\leq&\frac{1}{\pi}\Ep\left(\left[\rho(Y_i)\left\{K_h(Y_i-y_0)-b^*(\X_i,a_h^{*\dag},\Ep^\dag)\right\}\right]^2\right)
+\frac{1}{1-\pi}\Eq\left[\left\{b^*(\X_i,a_h^{*\dag},\Ep^\dag)-\delta_0\right\}^2\right]\n\\
&\leq&\frac{2}{\pi}\Ep\left\{\rho^2(Y_i)K_h^2(Y_i-y_0)\right\}+\frac{2}{\pi}\Ep\left\{\rho^2(Y_i)b^{*2}(\X_i,a_h^{*\dag},\Ep^\dag)\right\}\n\\
&&+\frac{2}{1-\pi}\Eq\left\{b^{*2}(\X_i,a_h^{*\dag},\Ep^\dag)\right\}+\frac{2\delta_0^2}{1-\pi}\n\\
&=&O\left\{\frac{N}{n}(h^{-1}+h^{-1})+\frac{N}{N-n}(h^{-1}+1)\right\}\n\\
&\leq&C_2\frac{N}{\min(n,N-n)}h^{-1}
\ee
for some constant $C_2>0$ uniformly for $i$,
because
\bse
\Ep\left\{\rho^2(Y_i)K_h^2(Y_i-y_0)\right\}
&=&\int\rho^2(y)K_h^2(y-y_0)\py(y)dy\\
&=&h^{-1}\int\rho^2(y_0+ht)K^2(t)\py(y_0+ht)dt\\
&=&h^{-1}\rho^2(y_0)\py(y_0)\int K^2(t)dt+O(1)\\
&=&O(h^{-1})
\ese
by Conditions \ref{con:density} and \ref{con:kernel},
and $\sup_\x|b^*(\x,a_h^{*\dag},\Ep^\dag)|=O(h^{-1/2})$ by
Conditions \ref{con:rho}, \ref{con:density}, and \eqref{eq:a}.

Now, let $t_1, \dots, t_{M_n}$ be arbitrary points on the
  support of $p_Y(y)$,
  and $M_n$ is either finite or diverging to infinity as
  $n\to\infty$, and $\wt q_Y(t_k)=\wt\delta_0$ at $y_0=t_k$. Then for
$\epsilon=C\{\min(n,N-n)h\}^{-1/2}\log M_n$ with some constant $C>0$, \eqref{eq:wttheta} leads to
\bse
&&\Pr\left(\max_{k=1,\dots,M_n}\left|\wt q_Y(t_k)-\Eq\{K_h(Y-t_k)\}\right|\geq\epsilon\right)\\
&\leq&M_n\Pr(|\wt\delta_0-\delta_0|\geq\epsilon)\\
&=&M_n\Pr\left\{\left|N^{-1}\sumI Z_i+o_p(n^{-1/2}h^{-1/2})\right|\geq\epsilon\right\}\\
&\leq&M_n\Pr\left\{\left|N^{-1}\sumI Z_i\right|\geq\epsilon+\left|o_p(n^{-1/2}h^{-1/2})\right|\right\}\\
&\leq&M_n\Pr\left(\left|N^{-1}\sumI Z_i\right|\geq2\epsilon\right)\\
&\leq&2M_n\exp\left(-\frac{2\epsilon^2}{2B\epsilon/3+V}\right)
\ese
by the Bernstein inequality where
\bse
B&=&C_1[n^{-1}h^{-1}+\{\min(n,N-n)\}^{-1}h^{-1/2}],\\
V&=&C_2\{\min(n,N-n)^{-1}h^{-1}\},
\ese
by \eqref{eq:bernstein1} and \eqref{eq:bernstein2}.
Also, $\Eq\{K_h(Y-t_k)\}=\qy(t_k)+O(h^m)$ by
Conditions \ref{con:density} and \ref{con:kernel}. These imply
\bse
\max_{k=1,\dots,M_n}\left|\wt
  q_Y(t_k)-q_Y(t_k)\right|=O(h^m)+O_p[\{\min(n,N-n)h\}^{-1/2}\log M_n].
\ese
On the other hand,
\bse
\max_{k=1,\dots,M_n}|\wh p_Y(t_k)-\py(t_k)|=O(h^m)+O_p\{(nh)^{-1/2}\log{M_n}\}
\ese
under Conditions \ref{con:density}-\ref{con:bandwidth}. Therefore,
\bse
\max_{k=1,\dots,M_n}\left|\wt\rho(t_k)-\rho(t_k)\right|
&=&\max_{k=1,\dots,M_n}\left|\frac{\wt q_Y(t_k)-q_Y(t_k)}{\wh p_Y(t_k)}+\qy(t_k)\left\{\frac{1}{\wh p_Y(t_k)}-\frac{1}{\py(t_k)}\right\}\right|\\
&=&O(h^m)+O_p[\{\min(n,N-n)h\}^{-1/2}\log M_n +(nh)^{-1/2}\log M_n]\\
&=&O(h^m)+O_p[\{\min(n,N-n)h\}^{-1/2}\log M_n],
\ese
since $\py$ and $\qy$ are bounded and bounded away from zero by Condition \ref{con:density}.

Now, let $t_1<\dots<t_{M(\delta_n)}$ be equally-spaced points
  on the support of $\rho(y)$, say $[a,b]$ under Condition
    \ref{con:density}, with grid spacing distance $\delta_n=C_\delta
  n^{m(1-m)/(2m+1)}$
  for some constant $C_\delta>0$. Let $y\in[a,b]$ and $t_k$ be the
  point closest to $y$ among $\{t_1,\dots,t_{M(\delta_n)}\}$, then
considering that $\wt\rho(\cdot)$ is a piecewise linear interpolation
of $\wt\rho(t_1),\dots,\wt\rho(t_{M(\delta_n)})$, we get
\bse
&&|\wt\rho(y)-\rho(y)|\\
&\leq&|\wt\rho(y)-\wt\rho(t_k)| + |\wt\rho(t_k)-\rho(t_k)| + |\rho(t_k)-\rho(y)|\\
&\leq&|\wt\rho(t_{k+1})-\wt\rho(t_k)| + |\wt\rho(t_k)-\rho(t_k)| + |\rho(t_k)-\rho(y)|\\
&\leq&|\wt\rho(t_{k+1})-\rho(t_{k+1})|+|\rho(t_{k+1})-\rho(t_k)|
+|\rho(t_k)-\wt\rho(t_k)| + |\wt\rho(t_k)-\rho(t_k)| + |\rho(t_k)-\rho(y)|\\
&\leq& 3\max_{k=1,\dots,M(\delta_n)} |\wt\rho(t_k)-\rho(t_k)| + 2L\delta_n,
\ese
where $L>0$ is the Lipschitz constant of $\rho(\cdot)$. Note that
\bse
n^{m(1-m)/(2m+1)}=o\{(nl^{2m})^{m/(2m+1)}\}=o(h^m)
\ese
under Condition \ref{con:bandwidth}.
Also note that $M(\delta_n)\asymp n^{m(m-1)/(2m+1)}$.  Therefore,
\bse
\|\wt\rho(y)-\rho(y)\|_\infty&=&
\sup_{y\in[a,b]}|\wt\rho(y)-\rho(y)|\\
&\leq&3\max_{k=1,\dots,M(\delta_n)} |\wt\rho(t_k)-\rho(t_k)| + 2L\delta_n\\
&=&O(h^m) + O_p[\{\min(n,N-n)h\}^{-1/2}\log\{M(\delta_n)\}] + o(h^{m})\\
&=&O(h^m) + O_p[\{\min(n,N-n)h\}^{-1/2}\log n].
\ese
\hfill$\blacksquare$


\section{Proof of Theorem \ref{th:thetaeff}}\label{supp:sec:proofthm3}
Let us define
\bse
\wt b(\x,a,\Ep)
&\equiv&\frac{\Ep\{a(Y)\wt\rho(Y)\mid\x\}}{\Ep\{\wt\rho^2(Y)+\pi/(1-\pi)\wt\rho(Y)\mid\x\}},\\
\wt\calI_{l,i}(a)(y)
&\equiv&\frac{r_i}{\pi}\wt K_l(y-y_i)\wt b(\x_i,a,\Ep),\\
\wh\calI(a)(y)
&\equiv&N^{-1}\sumI\frac{r_i}{\pi}\wt K_l(y-y_i)\wt b(\x_i,a,\wh\E_p),\\
d_{h,l,i}(y)
&\equiv&\wt\calI^{-1}\{v_{h,l,i}-\wt\calI_{l,i}(\wt a_h)\}(y).
\ese
Note that
\be
\E\{\wt b(\X,\wt a_h,\Ep)\mid y\}&=&\frac{\wt\calI(\wt a_h)(y)}{\py(y)}=\frac{v_h(y)}{\py(y)}=K_h(y-y_0).\label{eq:eb-eff}
\ee
We also define the $k$th Gateaux derivative of $g(\cdot,\mu)$ with respect to $\mu$ at $\mu_1$ in the direction $\mu_2$ as
\bse
\frac{\partial^k g(\cdot,\mu_1)}{\partial\mu^k}(\mu_2)
\equiv\frac{\partial^k g(\cdot,\mu)}{\partial\mu^k}(\mu_2)\Big|_{\mu=\mu_1}
\equiv\frac{\partial^k g(\cdot,\mu_1+h\mu_2)}{\partial h^k}\bigg|_{h=0}.
\ese
Then,
\be
&&\frac{\partial \wt b(\x,a,\Ep)}{\partial\Ep}(\wh\E_p-\Ep)\label{eq:b'form}\\
&=&\frac{(\wh\E_p-\Ep)\{a(Y)\wt\rho(Y)\mid\x\}}
{\Ep\{\wt\rho^2(Y)+\pi/(1-\pi)\wt\rho(Y)\mid\x\}}\n\\
&&-\frac{\Ep\{a(Y)\wt\rho(Y)\mid\x\}(\wh\E_p-\Ep)\{\wt\rho^2(Y)+\pi/(1-\pi)\wt\rho(Y)\mid\x\}}
{[\Ep\{\wt\rho^2(Y)+\pi/(1-\pi)\wt\rho(Y)\mid\x\}]^2}\n\\
&=&\wt b(\x,a,\Ep)\n\\
&&\times\left[\frac{(\wh\E_p-\Ep)\{a(Y)\wt\rho(Y)\mid\x\}}{\Ep\{a(Y)\wt\rho(Y)\mid\x\}}
-\frac{(\wh\E_p-\Ep)\{\wt\rho^2(Y)+\pi/(1-\pi)\wt\rho(Y)\mid\x\}}
{\Ep\{\wt\rho^2(Y)+\pi/(1-\pi)\wt\rho(Y)\mid\x\}}\right],\n\\
&&\frac{\partial^2\wt b(\x,a,\wt\E_p)}{\partial{\Ep}^2}(\wh\E_p-\Ep)\label{eq:b''form}\\
&=&\frac{-2(\wh\E_p-\Ep)\{\wt\rho^2(Y)+\pi/(1-\pi)\wt\rho(Y)\mid\x\}}
{[\wt\E_p\{\wt\rho^2(Y)+\pi/(1-\pi)\wt\rho(Y)\mid\x\}]^3}\n\\
&&\times\left[(\wh\E_p-\Ep)\{a(Y)\wt\rho(Y)\mid\x\}\wt\E_p\{\wt\rho^2(Y)+\pi/(1-\pi)\wt\rho(Y)\mid\x\}\right.\n\\
&&\left.-\wt\E_p\{a(Y)\wt\rho(Y)\mid\x\}(\wh\E_p-\Ep)\{\wt\rho^2(Y)+\pi/(1-\pi)\wt\rho(Y)\mid\x\}\right].\n
\ee

First, we derive the norms of some functions. We have
\be\label{eq:a-eff}
\|\wt a_h\|_2=\|\wt\calI^{-1}(v_h)\|_2=O(h^{-1/2}),
\ee
because $\wt\calI^{-1}$ is bounded by Lemma \ref{lem:i} and $\|v_h\|_2=O(h^{-1/2})$. In addition,
\bse
\|\E\{\wt\calI^{-1}(V_{h,l,i})\}-\wt\calI^{-1}(v_h)\|_2
=\|\wt\calI^{-1}\{\E(V_{h,l,i})-v_h\}\|_2
=O(h^{-1/2}l^m),
\ese
where the last equality holds by Lemma \ref{lem:i} and
$\|\E(V_{h,l,i})-v_h\|_2=O(h^{-1/2}l^m)$ under Conditions
\ref{con:density} and \ref{con:kernel}. Also,
\bse
\|\E\{(\wt\calI^{-1}\circ\wt\calI_{l,i})(\wt a_h)\}-\wt a_h\|_2
&=&\|\E\{(\wt\calI^{-1}\circ\wt\calI_{l,i})(\wt a_h)\}-\wt\calI^{-1}(v_h)\|_2\\
&=&\|\wt\calI^{-1}[\E\{\wt\calI_{l,i}(\wt a_h)\}-v_h]\|_2\\
&=&O\left[\left\|\E\left\{\frac{R}{\pi}\wt K_l(y-Y)\wt b(\X,\wt a_h,\Ep)\right\}-v_h\right\|_2\right]\\
&=&O\left[\|\py(y)\E\{\wt b(\X,\wt a_h,\Ep)\mid y\}+O(h^{-1/2}l^m)-v_h\|_2\right]\\
&=&O(h^{-1/2}l^m),
\ese
where the third equality holds by Lemma \ref{lem:i}, the fourth equality holds by \eqref{eq:a-eff}, Conditions \ref{con:density}, and \ref{con:kernel}, and the last equality holds by \eqref{eq:eb-eff}. Then,
\be\label{eq:edelta-eff}
\|\E(D_{h,l,i})\|_2
&=&\|\E\{\wt\calI^{-1}(V_{h,l,i})\}-\E\{(\wt\calI^{-1}\circ\wt\calI_{l,i})(\wt a_h)\}-\{\wt\calI^{-1}(v_h)-\wt a_h\}\|_2\n\\
&\leq&\|\E\{\wt\calI^{-1}(V_{h,l,i})\}-\wt\calI^{-1}(v_h)\|_2+\|\E\{(\wt\calI^{-1}\circ\wt\calI_{l,i})(\wt a_h)\}-\wt a_h\|_2\n\\
&=&O(h^{-1/2}l^m).
\ee
In addition,
\be\label{eq:deltaclt-eff}
&&\left\|N^{-1}\sumI d_{h,l,i}-\E(D_{h,l,i})\right\|_2\n\\
&=&\left\|N^{-1}\sumI\wt\calI^{-1}\{v_{h,l,i}-\wt\calI_{l,i}(\wt a_h)\}-\E\left[\wt\calI^{-1}\{V_{h,l,i}-\wt\calI_{l,i}(\wt a_h)\}\right]\right\|_2\n\\
&=&\left\|\wt\calI^{-1}\left[N^{-1}\sumI v_{h,l,i}-\E(V_{h,l,i})-N^{-1}\sumI\wt\calI_{l,i}(\wt a_h)+\E\{\wt\calI_{l,i}(\wt a_h)\}\right]\right\|_2\n\\
&=&O\left[\left\|N^{-1}\sumI v_{h,l,i}-\E(V_{h,l,i})\right\|_2+\left\|N^{-1}\sumI\wt\calI_{l,i}(\wt a_h)-\E\{\wt\calI_{l,i}(\wt a_h)\}\right\|_2\right]\n\\
&=&O_p\{h^{-1/2}(nl)^{-1/2}\},
\ee
where the second equality holds since $\wt\calI^{-1}$ is linear by
Lemma \ref{lem:i}, the third equality holds since $\calI^{-1}$ is
bounded by Lemma \ref{lem:i}, and the last equality holds by
\eqref{eq:a-eff} and Condition \ref{con:kernel}. On the other hand,
for $a(y)$ such that $\|a\|_2<\infty$, \eqref{eq:b'form} leads to
\be\label{eq:b'bounded-eff}
&&\left|\frac{\partial \wt b(\x,a,\Ep)}{\partial\Ep}(\wh\E_p-\Ep)\right|\n\\
&=&\left|\wt b(\x,a,\Ep)\right|\n\\
&&\times\left|\frac{(\wh\E_p-\Ep)\{a(Y)\wt\rho(Y)\mid\x\}}{\Ep\{a(Y)\wt\rho(Y)\mid\x\}}
-\frac{(\wh\E_p-\Ep)\{\wt\rho^2(Y)+\pi/(1-\pi)\wt\rho(Y)\mid\x\}}
{\Ep\{\wt\rho^2(Y)+\pi/(1-\pi)\wt\rho(Y)\mid\x\}}\right|\n\\
&\leq&\frac{\|a\|_2\|\wt\rho(\cdot)\pyx(\cdot,\x)\|_2}{\Ep\{\wt\rho^2(Y)+\pi/(1-\pi)\wt\rho(Y)\mid\x\}}\n\\
&&\times\left|\frac{(\wh\E_p-\Ep)\{a(Y)\wt\rho(Y)\mid\x\}}{\Ep\{a(Y)\wt\rho(Y)\mid\x\}}
-\frac{(\wh\E_p-\Ep)\{\wt\rho^2(Y)+\pi/(1-\pi)\wt\rho(Y)\mid\x\}}
{\Ep\{\wt\rho^2(Y)+\pi/(1-\pi)\wt\rho(Y)\mid\x\}}\right|\n\\
&=&o_p(n^{-1/4}\|a\|_2)
\ee
uniformly in $\x$ by Condition \ref{con:density}, because $\wt\rho(y)$
is bounded away from zero by Condition \ref{con:rho}, $\wt\rho(y)$ is
bounded above since its derivative is bounded on a compact support by
Conditions \ref{con:rho} and \ref{con:density}, and
$\wt\rho(\cdot)\pyx^\dag(\cdot,\x,\bb)$ is square integrable by
Condition \ref{con:pyx}, and
$|(\wh\E_p-\Ep)\{a(Y)\mid\x\}|=o_p(n^{-1/4}\|a\|_2)$. Similarly, for
$\wt\E_p=\Ep+\alpha(\wh\E_p-\Ep)$ with $\alpha\in(0,1)$,
\eqref{eq:b''form} leads to
\bse
\left|\frac{\partial^2\wt b(\x,a,\wt\E_p)}{\partial{\Ep}^2}(\wh\E_p-\Ep)\right|=o_p(n^{-1/2}\|a\|_2)
\ese
uniformly in $\x$. Thus,
\be
\wt b(\x,a,\wh\E_p)-\wt b(\x,a,\Ep)
&=&\frac{\partial \wt b(\x,a,\Ep)}{\partial\Ep}(\wh\E_p-\Ep)
+\frac{1}{2}\frac{\partial^2 \wt b(\x,a,\wt\E_p)}{\partial{\Ep}^2}(\wh\E_p-\Ep)\n\\
&=&\frac{\partial \wt b(\x,a,\Ep)}{\partial\Ep}(\wh\E_p-\Ep)
+o_p(n^{-1/2}\|a\|_2)\label{eq:b'-eff}\\
&=&o_p(n^{-1/4}\|a\|_2)\n
\ee
uniformly in $\x$. Hence,
\be\label{eq:ihat-eff}
\|(\wh\calI-\wt\calI)(a)\|_2
&=&\left\|N^{-1}\sumI\frac{r_i}{\pi}\wt K_l(y-y_i)\wt b(\x_i,a,\wh\E_p)-\py(y)\E\{\wt b(\x,a,\Ep)\mid y\}\right\|_2\n\\
&\leq&\left\|N^{-1}\sumI\frac{r_i}{\pi}\wt K_l(y-y_i)\left\{\wt b(\x_i,a,\wh\E_p)-\wt b(\x_i,a,\Ep)\right\}\right\|_2\n\\
&&+\left\|N^{-1}\sumI\frac{r_i}{\pi}\wt K_l(y-y_i)\wt b(\x_i,a,\Ep)-\py(y)\E\{\wt b(\x,a,\Ep)\mid y\}\right\|_2\n\\
&=&o_p(n^{-1/4}\|a\|_2)+O_p\left[\{(nl)^{-1/2}+l^m\}\|a\|_2\right]=o_p(n^{-1/4}\|a\|_2)
\ee
under Conditions \ref{con:u}-\ref{con:bandwidth}. Similarly,
\be\label{eq:vhat-eff}
\|\wh v-v_h\|_2= O_p\left\{(nl)^{-1/2}+l^m\right\}O(h^{-1/2})=o_p(n^{-1/4}h^{-1/2}).
\ee
Furthermore,
\be\label{eq:ihat2-eff}
&&\wh\calI(\wt a_h)(y)\n\\
&=&N^{-1}\sumI\frac{r_i}{\pi}\wt K_l(y-y_i)\left\{\wt b(\x_i,\wt a_h,\Ep)
+\frac{\partial \wt b(\x_i,\wt a_h,\Ep)}{\partial\Ep}
(\wh\E_p-\Ep)+o_p(n^{-1/2}h^{-1/2})\right\}\n\\
&=&N^{-1}\sumI\wt\calI_{l,i}(\wt a_h)(y)
+\py(y)\frac{\partial\E\{\wt b(\X,\wt a_h,\Ep)\mid y\}}{\partial\Ep}
(\wh\E_p-\Ep)\n\\
&&+o_p[\{(nl)^{-1/2}+l^m\}n^{-1/4}h^{-1/2}]+o_p(n^{-1/2}h^{-1/2})\n\\
&=& N^{-1}\sumI\wt\calI_{l,i}(\wt a_h)(y)
+o_p(n^{-1/2}h^{-1/2}),
\ee
where the first equality holds by \eqref{eq:a-eff} and
\eqref{eq:b'-eff}, the second equality holds by \eqref{eq:a-eff},
\eqref{eq:b'bounded-eff}, and Conditions \ref{con:u}-\ref{con:kernel},
and the last equality holds by \eqref{eq:eb-eff} and Condition
\ref{con:bandwidth}.

Now, we derive the asymptotic form of $\wt b(\x,\wh a,\wh\E_p)$. $\wh a$ can be written as, since $\wt\calI^{-1}$ is linear and bounded by Lemma \ref{lem:i},
\be\label{eq:ahat-eff}
\wh a(y)
&=&\wh\calI^{-1}(\wh v)(y)\n\\
&=&\{\wt\calI+(\wh\calI-\wt\calI)\}^{-1}(\wh v)(y)\n\\
&=&\{\wt\calI^{-1}-\wt\calI^{-1}\circ(\wh\calI-\wt\calI)\circ\wt\calI^{-1}\}\{v_h+(\wh v-v_h)\}(y)+o_p(n^{-1/2}h^{-1/2})\n\\
&=&\wt a_h(y)+\wt\calI^{-1}(\wh v-v_h)(y)-\{\wt\calI^{-1}\circ(\wh\calI-\wt\calI)\}(\wt a_h)(y)+o_p(n^{-1/2}h^{-1/2})\n\\
&=&\wt a_h(y)+\wt\calI^{-1}\{\wh v-\wh\calI(\wt a_h)\}(y)+o_p(n^{-1/2}h^{-1/2})\n\\
&=&\wt a_h(y)+N^{-1}\sumI d_{h,l,i}(y)+o_p(n^{-1/2}h^{-1/2})
\ee
uniformly in $y$ by Condition \ref{con:density}. Here, the third equality holds by \eqref{eq:ihat-eff} and
\bse
\|\wh v\|_2\leq\|v_h\|_2+\|\wh v-v_h\|_2=O(h^{-1/2})+o_p(n^{-1/4}h^{-1/2})=O_p(h^{-1/2})
\ese
by \eqref{eq:vhat-eff}, the fourth equality holds by $\wt
a_h(y)=\wt\calI^{-1}(v_h)(y)$, \eqref{eq:ihat-eff}, and
\eqref{eq:vhat-eff}, and the last equality holds by
\eqref{eq:ihat2-eff}. Then
\be\label{eq:ahat-a-eff}
\|\wh a-\wt a_h\|_2=O(h^{-1/2}l^m)+O_p\{h^{-1/2}(nl)^{-1/2}\}+o_p(n^{-1/2}h^{-1/2})=o_p(n^{-1/4}h^{-1/2})
\ee
by \eqref{eq:edelta-eff}, \eqref{eq:deltaclt-eff}, \eqref{eq:ahat-eff}, and Condition \ref{con:bandwidth}, and
\be\label{eq:ahat2-eff}
\|\wh a\|_2\leq\|\wt a_h\|_2+\|\wh a-\wt a_h\|_2=O_p(h^{-1/2})
\ee
by \eqref{eq:a-eff}. Further, \eqref{eq:b'form} leads to
\be\label{eq:b'2-eff}
&&\left|\frac{\partial \wt b(\x,\wh a,\Ep)}{\partial\Ep}(\wh\E_p-\Ep)
-\frac{\partial \wt b(\x,\wt a_h,\Ep)}{\partial\Ep}(\wh\E_p-\Ep)\right|\n\\
&=&\left|\frac{(\wh\E_p-\Ep)\{(\wh a-\wt a_h)(Y)\wt\rho(Y)\mid\x\}}
{\Ep\{\wt\rho^2(Y)+\pi/(1-\pi)\wt\rho(Y)\mid\x\}}\right.\n\\
&&\left.-\frac{\Ep\{(\wh a-\wt a_h)(Y)\wt\rho(Y)\mid\x\}
(\wh\E_p-\Ep)\{\wt\rho^2(Y)+\pi/(1-\pi)\wt\rho(Y)\mid\x\}}
{[\Ep\{\wt\rho^2(Y)+\pi/(1-\pi)\wt\rho(Y)\mid\x\}]^2}\right|\n\\
&\leq&\frac{|(\wh\E_p-\Ep)\{(\wh a-\wt a_h)(Y)\wt\rho(Y)\mid\x\}|}
{\Ep\{\wt\rho^2(Y)+\pi/(1-\pi)\wt\rho(Y)\mid\x\}}\n\\
&&+\frac{\|\wh a-\wt a_h\|_2\|\wt\rho(\cdot)\pyx(\cdot,\x)\|_2
|(\wh\E_p-\Ep)\{\wt\rho^2(Y)+\pi/(1-\pi)\wt\rho(Y)\mid\x\}|}
{[\Ep\{\wt\rho^2(Y)+\pi/(1-\pi)\wt\rho(Y)\mid\x\}]^2}\n\\
&=&o_p(n^{-1/4} n^{-1/4}h^{-1/2})+o_p(n^{-1/4}h^{-1/2} n^{-1/4})\n\\
&=&o_p(n^{-1/2}h^{-1/2}),
\ee
because $\wt\rho(y)$ is bounded away from zero and above by Conditions
\ref{con:rho} and \ref{con:density},
$|(\wh\E_p-\Ep)\{a(Y)\mid\x\}|=o_p(n^{-1/4}\|a\|_2)$, and $\|\wh a-\wt
a_h\|_2=o_p(n^{-1/4}h^{-1/2})$ by \eqref{eq:ahat-a-eff}. Hence,
\be\label{eq:bhat-eff}
&&\wt b(\x,\wh a,\wh\E_p)\\
&=&\wt b(\x,\wh a,\Ep)
+\frac{\partial \wt b(\x,\wh a,\Ep)}{\partial\Ep}(\wh\E_p-\Ep)+o_p(n^{-1/2}h^{-1/2})\n\\
&=&\wt b(\x,\wt a_h,\Ep)+\wt b(\x,\wh a-\wt a_h,\Ep)
+\frac{\partial \wt b(\x,\wt a_h,\Ep)}{\partial\Ep}(\wh\E_p-\Ep)+o_p(n^{-1/2}h^{-1/2})\n\\
&=&\wt b(\x,\wt a_h,\Ep)+N^{-1}\sumI \wt b(\x,d_{h,l,i},\Ep)
+\frac{\partial \wt b(\x,\wt a_h,\Ep)}{\partial\Ep}(\wh\E_p-\Ep)+o_p(n^{-1/2}h^{-1/2})\n
\ee
uniformly in $\x$ by Condition \ref{con:density}, where the first
equality holds by \eqref{eq:b'-eff} and \eqref{eq:ahat2-eff}, the
second equality holds by \eqref{eq:b'2-eff}, and the last equality
holds by \eqref{eq:ahat-eff}.

Now, we analyze $\wh\delta_0$. From the definition of $\wh\delta_0$,
\be\label{eq:theta-eff}
\wh\delta_0-\delta_0
&=&N^{-1}\sumI\left[\frac{r_i}{\pi}\wt\rho(y_i)\{K_h(y_i-y_0)-\wt b(\x_i,\wh a,\wh\E_p)\}
+\frac{1-r_i}{1-\pi}\{\wt b(\x_i,\wh a,\wh\E_p)-\delta_0\}\right]\n\\
&=&-N^{-1}\sumI\wt\phi\eff(r_i,r_iy_i,\x_i)\n\\
&&-N^{-1}\sumI\left\{\frac{r_i}{\pi}\wt\rho(y_i)-\frac{1-r_i}{1-\pi}\right\}
\{\wt b(\x_i,\wh a,\wh\E_p)-\wt b(\x_i,\wt a_h,\Ep)\}\n\\
&=&-N^{-1}\sumI\wt\phi\eff(r_i,r_iy_i,\x_i)-T_1-T_2+o_p(n^{-1/2}h^{-1/2}),
\ee
where $\wt\phi\eff(r,ry,\x)$ is $\phi\eff(r,ry,\x)$ with $\rho(y)$ replaced by $\wt\rho(y)$,
\bse
T_1&\equiv&N^{-2}\sumI\sumJ\left\{\frac{r_i}{\pi}\wt\rho(y_i)-\frac{1-r_i}{1-\pi}\right\}\wt b(\x_i,d_{h,l,j},\Ep),\\
T_2&\equiv&N^{-1}\sumI\left\{\frac{r_i}{\pi}\wt\rho(y_i)-\frac{1-r_i}{1-\pi}\right\}
\frac{\partial \wt b(\x_i,\wt a_h,\Ep)}{\partial\Ep}(\wh\E_p-\Ep).
\ese
By the property of the U-statistic, $T_1$ can be written as
\be\label{eq:t1-eff}
&&T_1\\
&=&N^{-1}\sumI
\left\{\frac{r_i}{\pi}\wt\rho(y_i)-\frac{1-r_i}{1-\pi}\right\}
\E\{\wt b(\x_i,D_{h,l,j},\Ep)\mid r_i,r_iy_i,\x_i\}\n\\
&&+N^{-1}\sumJ
\E\left[\left\{\frac{R_i}{\pi}\wt\rho(Y_i)-\frac{1-R_i}{1-\pi}\right\}
\wt b(\X_i,d_{h,l,j},\Ep)\mid r_j,r_jy_j,\x_j\right]\n\\
&&-N^{-1/2}\E\left[\left\{\frac{R_i}{\pi}\wt\rho(Y_i)-\frac{1-R_i}{1-\pi}\right\}
\wt b(\X_i,D_{h,l,j},\Ep)\right]+O_p(N^{-1}h^{-1/2})\n\\
&=&N^{-1}\sumJ\E\left[\left\{\frac{R_i}{\pi}\wt\rho(Y_i)-\frac{1-R_i}{1-\pi}\right\}
\wt b(\X_i,d_{h,l,j},\Ep)\mid r_j,r_jy_j,\x_j\right]\n\\
&&+O_p(h^{-1/2}l^m+N^{-1}h^{-1/2})\n\\
&=&N^{-1}\sumJ\int\left[\wt\rho(y)\E\{\wt b(\X,d_{h,l,j},\Ep)\mid y\}\py(y)
-\E\{\wt b(\X,d_{h,l,j},\Ep)\mid y\}\qy(y)\right]dy\n\\
&&+o_p(n^{-1/2}h^{-1/2})\n\\
&=&N^{-1}\sumJ\int\left\{\wt\rho(y)-\rho(y)\right\}\wt\calI(d_{h,l,j})(y)dy+o_p(n^{-1/2}h^{-1/2})\n\\
&=&N^{-1}\sumJ\int\left\{\wt\rho(y)-\rho(y)\right\}\{v_{h,l,j}(y)-\wt\calI_{l,j}(\wt a_h)(y)\}dy+o_p(n^{-1/2}h^{-1/2})\n\\
&=&N^{-1}\sumJ\frac{r_j}{\pi}\int \wt K_l(y-y_j)\left\{\wt\rho(y)-\rho(y)\right\}\{K_h(y-y_0)-\wt b(\x_j,\wt a_h,\Ep)\}dy+o_p(n^{-1/2}h^{-1/2}),\n
\ee
where the second equality holds because \eqref{eq:edelta-eff} leads to
\bse
\E\{\wt b(\x_i,D_{h,l,j},\Ep)\mid r_i,r_iy_i,\x_i\}=O(h^{-1/2}l^m),
\ese
and the third equality holds by Condition \ref{con:bandwidth}. Further,
\be\label{eq:t12-eff}
&&\int \wt K_l(y-y_j)\left\{\wt\rho(y)-\rho(y)\right\}\{K_h(y-y_0)-\wt b(\x_j,\wt a_h,\Ep)\}dy\n\\
&=&\int \wt K(t)\{\wt\rho(y_j+lt)-\rho(y_j+lt)\}\{K_h(y_j+lt-y_0)-\wt b(\x_j,\wt a_h,\Ep)\}dt\n\\
&=&\{\wt\rho(y_j)-\rho(y_j)\}\{K_h(y_j-y_0)-\wt b(\x_j,\wt a_h,\Ep)\}+
O(h^{-m-1}l^m)\n\\
&=&\{\wt\rho(y_j)-\rho(y_j)\}\{K_h(y_j-y_0)-\wt b(\x_j,\wt a_h,\Ep)\}+o(n^{-1/2}h^{-1/2})
\ee
under Conditions \ref{con:rho}, \ref{con:kernel}, and
\ref{con:bandwidth}. On the other hand, $T_2$ can be written as
\be\label{eq:t2-eff}
T_2
&=&\E\left[\left\{\frac{R}{\pi}\wt\rho(Y)-\frac{1-R}{1-\pi}\right\}
\frac{\partial \wt b(\X,\wt a_h,\Ep)}{\partial\Ep}(\wh\E_p-\Ep)\right]
+O_p(n^{-1/2})o_p(n^{-1/4}h^{-1/2})\n\\
&=&\frac{\partial}{\partial\Ep}
\E\left[\left\{\frac{R}{\pi}\wt\rho(Y)-\frac{1-R}{1-\pi}\right\}K_h(Y-y_0)\right]
(\wh\E_p-\Ep)+o_p(n^{-1/2}h^{-1/2})\n\\
&=&o_p(n^{-1/2}h^{-1/2}),
\ee
where the first equality holds by \eqref{eq:a-eff} and \eqref{eq:b'bounded-eff}, and the second equality holds by \eqref{eq:eb-eff}.
Therefore, \eqref{eq:theta-eff}, \eqref{eq:t1-eff}, \eqref{eq:t12-eff}
and \eqref{eq:t2-eff} lead to
\be\label{eq:wttheta-eff}
&&\wh\delta_0-\delta_0\n\\
&=&-N^{-1}\sumI\wt\phi\eff(r_i,r_iy_i,\x_i)\n\\
&&-N^{-1}\sumI\frac{r_i}{\pi}\{\wt\rho(y_i)-\rho(y_i)\}\{K_h(y_i-y_0)-\wt b(\x_i,\wt a_h,\Ep)\}+o_p(n^{-1/2}h^{-1/2})\n\\
&=&N^{-1}\sumI\frac{r_i}{\pi}\rho(y_i)\left\{K_h(y_i-y_0)-\wt b(\x_i,\wt a_h,\Ep)\right\}\n\\
&&+N^{-1}\sumI\frac{1-r_i}{1-\pi}\left\{\wt b(\x_i,\wt a_h,\Ep)-\delta_0\right\}+o_p(n^{-1/2}h^{-1/2})\n\\
&=&N^{-1}\sumI\frac{r_i}{\pi}\rho(y_i)K_h(y_i-y_0)-N^{-1}\sumI\left\{\frac{r_i}{\pi}\rho(y_i)-\frac{1-r_i}{1-\pi}\right\}\wt b(\x_i,\wt a_h,\Ep)\n\\
&&-\delta_0+o_p(n^{-1/2}h^{-1/2}).
\ee

Now, for $a(y)$ such that $\|a\|_2<\infty$,
\be\label{eq:wtcalI-eff}
\|(\wt\calI-\calI)(a)\|_2
&=&\left[\int\left\{\int a(t)(\wt u-u)(t,y)dt\right\}^2dy\right]^{1/2}\n\\
&\leq&\left[\int\|a\|_2^2\|(\wt u-u)(\cdot,y)\|_2^2dy\right]^{1/2}\n\\
&=&o_p(\|a\|_2)
\ee
since $\sup_y|\wt\rho(y)-\rho(y)|=o_p(1)$. Recall that $a_h(y)$ in $\bphi\eff(r,ry,\x)$ satisfies $\calI(a_h)(y)=v_h(y)$, and $\wt a_h(y)$ satisfies $\wt\calI(\wt a_h)(y)=v_h(y)$. Also, recall Lemma \ref{lem:i} to have that both $\calI$ and $\wt\calI$ are linear, invertible, bounded, and their inverses are also bounded. Then,
\bse
\wt a_h(y)
&=&\wt\calI^{-1}(v_h)(y)\n\\
&=&\{\calI+(\wt\calI-\calI)\}^{-1}(v_h)(y)\n\\
&=&\{\calI^{-1}-\calI^{-1}\circ(\wt\calI-\calI)\circ\calI^{-1}\}(v_h)(y)+o_p(h^{-1/2})\n\\
&=&a_h(y)-\{\calI^{-1}\circ(\wt\calI-\calI)\}(a_h)(y)+o_p(h^{-1/2})\n\\
&=&a_h(y)+o_p(h^{-1/2}),
\ese
where the third equality holds by \eqref{eq:wtcalI-eff} and $\|v_h\|_2=O(h^{-1/2})$, and the last equality holds by \eqref{eq:wtcalI-eff} and $\|a_h\|_2=\|\calI^{-1}(v_h)\|_2=O(h^{-1/2})$. This holds uniformly for $y$ by Condition \ref{con:density}. This and $\sup_y|\wt\rho(y)-\rho(y)|=o_p(1)$ further lead to
\bse
\wt b(\x,\wt a_h,\Ep)=b(\x,a_h,\Ep)+o_p(h^{-1/2})
\ese
where
\bse
b(\x,a,\Ep)=\frac{\Ep\{a(Y)\rho(Y)\mid\x\}}{\Ep\{\rho^2(Y)+\pi/(1-\pi)\rho(Y)\mid\x\}},
\ese
and this holds uniformly for $\x$ by Condition \ref{con:density}. Then,
\bse
&&N^{-1}\sumI\left\{\frac{r_i}{\pi}\rho(y_i)-\frac{1-r_i}{1-\pi}\right\}\wt b(\x_i,\wt a_h,\Ep)\\
&=&N^{-1}\sumI\left\{\frac{r_i}{\pi}\rho(y_i)-\frac{1-r_i}{1-\pi}\right\}b(\x_i,a_h,\Ep)\\
&&+\E\left[\left\{\frac{R}{\pi}\rho(Y)-\frac{1-R}{1-\pi}\right\}\E\left\{\wt b(\X,\wt a_h,\Ep)-b(\X,a_h,\Ep)\mid Y\right\}\right]+o_p(n^{-1/2}h^{-1/2})\\
&=&N^{-1}\sumI\left\{\frac{r_i}{\pi}\rho(y_i)-\frac{1-r_i}{1-\pi}\right\}b(\x_i,a_h,\Ep)+o_p(n^{-1/2}h^{-1/2}),
\ese
where the second equality holds since $\rho(y)$ is bounded under Condition \ref{con:density} and $\wt b(\x,\wt a_h,\Ep)-b(\x,a_h,\Ep)=o_p(h^{-1/2})$, and the last equality holds since $\E\{\wt b(\X,a_h,\Ep)-b(\X,a_h,\Ep)\mid y\}=K_h(y-y_0)-K_h(y-y_0)=0$.
This fact and \eqref{eq:wttheta-eff} lead to
\bse
\wh\delta_0-\delta_0
&=&N^{-1}\sumI\frac{r_i}{\pi}\rho(y_i)K_h(y_i-y_0)-N^{-1}\sumI\left\{\frac{r_i}{\pi}\rho(y_i)-\frac{1-r_i}{1-\pi}\right\}b(\x_i,a_h,\Ep)\\
&&-\delta_0+o_p(n^{-1/2}h^{-1/2})\\
&=&N^{-1}\sumI\frac{r_i}{\pi}\rho(y_i)\left\{K_h(y_i-y_0)-b(\x_i,a_h,\Ep)\right\}\\
&&+N^{-1}\sumI\frac{1-r_i}{1-\pi}\left\{b(\x_i,a_h,\Ep)-\delta_0\right\}+o_p(n^{-1/2}h^{-1/2}).
\ese
Furthermore,
\bse
\delta_0
&=&\int K_h(y-y_0)\qy(y)dy\\
&=&\int K(t)\qy(y_0+ht)dt\\
&=&\qy(y_0)+\qy^{(m)}(y_0)\frac{\int t^mK(t)dt}{m!}h^m+O(h^{m+1})
\ese
under Conditions \ref{con:density}-\ref{con:bandwidth}, and
\bse
\var(\wh\delta_0-\delta_0)
&=&N^{-1}\E\left(\left[\frac{R_i}{\pi}\rho(Y_i)\left\{K_h(Y_i-y_0)-b(\X_i,a_h,\Ep)\right\}\right]^2\right)\\
&&+N^{-1}\E\left(\left[\frac{1-R_i}{1-\pi}\left\{b(\X_i,a_h,\Ep)-\delta_0\right\}\right]^2\right)\\
&&+ o[\{\min(n,N-n)\}^{-1/2}n^{-1/2}h^{-1}]\\
&=&n^{-1}\E\left(\frac{R_i}{\pi}\left[\rho(Y_i)\left\{K_h(Y_i-y_0)-b(\X_i,a_h,\Ep)\right\}\right]^2\right)\\
&&+(N-n)^{-1}\E\left[\frac{1-R_i}{1-\pi}\left\{b(\X_i,a_h,\Ep)-\delta_0\right\}^2\right]\\
&&+ o[\{\min(n,N-n)\}^{-1/2}n^{-1/2}h^{-1}]\\
&=&n^{-1}\Ep\left(\left[\rho(Y_i)\left\{K_h(Y_i-y_0)-b(\X_i,a_h,\Ep)\right\}\right]^2\right)\\
&&+(N-n)^{-1}\Eq\left[\left\{b(\X_i,a_h,\Ep)-\delta_0\right\}^2\right]\\
&&+ o[\{\min(n,N-n)\}^{-1/2}n^{-1/2}h^{-1}].
\ese
Because the variance of  $\wh\delta_0$ is the variance of the
  efficient influence function of estimating $E_q\{K_h(Y-y)\}$, hence
  $\wh\delta_0$ is efficient.  Further at any fixed $K(\cdot), h$,  $\wh\delta_0$ is
  an efficient estimator of $q_Y(y)$.
\hfill$\blacksquare$


\section{Proof of Theorem \ref{th:general-eff}}\label{supp:sec:proofthm4}
We recall the last two displays in the proof of Theorem
S.2 in the supplement of \cite{lee2024doubly}. These equations with
$\rho^*=\wt\rho$ lead to
\bse
\0&=&N^{-1/2}\sumI\left[\sqrt\pi\wt\bphi\eff(r_i,r_iy_i,\x_i,\bt)+\frac{r_i}{\sqrt{\pi}}\wt\A(\bt)\{\wt\b(\x_i,\bt)-\U(y_i,\x_i,\bt)\}\{\wt\rho(y_i)-\rho(y_i)\}\right]\\
&&+\left\{\wt\A(\bt)\A(\bt)^{-1}+o_p(1)\right\}\sqrt{n}(\wh\bt-\bt)+o_p(1),
\ese
where
\bse
\wt\bphi\eff(r,ry,\x,\bt)&=&\frac{r}{\pi}\wt\rho(y)\wt\A(\bt)\{\U(y,\x,\bt)-\wt\b(\x,\bt)\}+\frac{1-r}{1-\pi}\wt\A(\bt)\wt\b(\x,\bt),\\
\wt\A(\bt)&=&\left(\E_p\left[\wt\rho(Y)\E\{\partial\U(Y,\X,\bt)/\partial\bt\trans\mid Y\}\right]\right)^{-1},\\
\wt\b(\x,\bt)&=&\frac{\E_p\{\U(Y,\x,\bt)\wt\rho^2(Y)+\wt\a(Y,\bt)\wt\rho(Y)\mid\x\}}{\E_p\{\wt\rho^2(Y)+\pi/(1-\pi)\wt\rho(Y)\mid\x\}},
\ese
and $\wt\a(y,\bt)$ satisfies
\bse
\E\left[\frac{\E_p\{\U(Y,\X,\bt)\wt\rho^2(Y)+\wt\a(Y,\bt)\wt\rho(Y)\mid\X\}}{\E_p\{\wt\rho^2(Y)+\pi/(1-\pi)\wt\rho(Y)\mid\X\}}\mid y\right]=\E\{\U(y,\X,\bt)\mid y\}.
\ese
This leads to
\be\label{eq:whbt}
&&-\left\{\A(\bt)^{-1}+o_p(1)\right\}\sqrt{n}(\wh\bt-\bt)+o_p(1)\\
&=&N^{-1/2}\sumI\sqrt\pi\left[\wt\A(\bt)^{-1}\wt\bphi\eff(r_i,r_iy_i,\x_i,\bt)+\frac{r_i}{\pi}\{\wt\b(\x_i,\bt)-\U(y_i,\x_i,\bt)\}\{\wt\rho(y_i)-\rho(y_i)\}\right]\n\\
&=&N^{-1/2}\sumI\sqrt\pi\left[\frac{r_i}{\pi}\wt\rho(y_i)\{\U(y_i,\x_i,\bt)-\wt\b(\x_i,\bt)\}+\frac{1-r_i}{1-\pi}\wt\b(\x_i,\bt)\right.\n\\
&&\left.+\frac{r_i}{\pi}\{\rho(y_i)-\wt\rho(y_i)\}\{\U(y_i,\x_i,\bt)-\wt\b(\x_i,\bt)\}\right]\n\\
&=&N^{-1/2}\sumI\sqrt{\pi}\left[\frac{r_i}{\pi}\rho(y_i)\{\U(y_i,\x_i,\bt)-\wt\b(\x_i,\bt)\}+\frac{1-r_i}{1-\pi}\wt\b(\x_i,\bt)\right]\n\\
&=&\sqrt{n}N^{-1}\sumI\frac{r_i}{\pi}\rho(y_i)\U(y_i,\x_i,\bt)-\sqrt{n}N^{-1}\sumI\left\{\frac{r_i}{\pi}\rho(y_i)-\frac{1-r_i}{1-\pi}\right\}\wt\b(\x_i,\bt).\n
\ee
Let
\bse
\calI(\a)(y,\bt)&\equiv&\py(y)\E\left[\frac{\E_p\{\a(Y,\bt)\rho(Y)\mid\X\}}{\E_p\{\rho^2(Y)+\pi/(1-\pi)\rho(Y)\mid\X\}}\mid y\right],\\
\v(y,\bt)&\equiv&\py(y)\E\left[\U(y,\X,\bt)-\frac{\E_p\{\U(Y,\X,\bt)\rho^2(Y)\mid\X\}}{\E_p\{\rho^2(Y)+\pi/(1-\pi)\rho(Y)\mid\X\}}\mid y\right],
\ese
and $\wt\calI(\a)(y,\bt),\wt\v(y,\bt)$ be $\calI(\a)(y,\bt),\v(y,\bt)$
with $\rho$ replaced by $\wt\rho$. For
$\a(y,\bt)=\{a_1(y,\bt),\dots,a_p(y,\bt)\}\trans$, let
$\|\a\|_\infty=\max_{k=1,\dots,p}\sup_{y,\bt}|a_k(y,\bt)|$. Then for
any $\a(y,\bt)$ with $\|\a\|_\infty<\infty$,
\be\label{eq:wtcalI}
\|(\wt\calI-\calI)(\a)\|_\infty=o_p(1)
\ee
since $\sup_y|\wt\rho(y)-\rho(y)|=o_p(1)$, and the support of $Y$ and
the parameter space of $\bt$ are compact under the regularity
conditions. Similarly,
\be\label{eq:wtv}
\|\wt\v-\v\|_\infty=o_p(1).
\ee
Note that $\a(y,\bt)$ in $\bphi\eff(r,ry,\x,\bt)$ satisfies
$\calI(\a)(y,\bt)=\v(y,\bt)$, and $\wt\a(y,\bt)$ satisfies
$\wt\calI(\wt\a)(y,\bt)=\wt\v(y,\bt)$. Also, we recall Lemma
  S.1 in the supplement of \cite{lee2024doubly} to have that both
  $\calI$ and $\wt\calI$ are linear, invertible, bounded, and their
  inverses are also bounded. Then,
\bse
\wt\a(y,\bt)
&=&\wt\calI^{-1}(\wt\v)(y,\bt)\n\\
&=&\{\calI+(\wt\calI-\calI)\}^{-1}\{\v+(\wt\v-\v)\}(y,\bt)\n\\
&=&\{\calI^{-1}-\calI^{-1}\circ(\wt\calI-\calI)\circ\calI^{-1}\}\{\v+(\wt\v-\v)\}(y,\bt)+o_p(1)\n\\
&=&\a(y,\bt)+\calI^{-1}(\wt\v-\v)(y,\bt)-\{\calI^{-1}\circ(\wt\calI-\calI)\}(\a)(y,\bt)+o_p(1)\n\\
&=&\a(y,\bt)+o_p(1),
\ese
where the third equality holds by \eqref{eq:wtcalI}, and the fourth
and last equality hold by \eqref{eq:wtcalI} and \eqref{eq:wtv}.
  This holds uniformly for $(y,\bt)$ by the compactness of the support
  of $Y$ and the parameter space of $\bt$ under the regularity
  conditions. This and $\sup_y|\wt\rho(y)-\rho(y)|=o_p(1)$ further
  lead to
\bse
\wt\b(\x,\bt)=\b(\x,\bt)+o_p(1)
\ese
where
\bse
\b(\x,\bt)=\frac{\E_p\{\U(Y,\x,\bt)\rho^2(Y)+\a(Y,\bt)\rho(Y)\mid\x\}}{\E_p\{\rho^2(Y)+\pi/(1-\pi)\rho(Y)\mid\x\}},
\ese
and this holds uniformly for $(\x,\bt)$ by the compactness of the
support of $\X$ and the parameter space of $\bt$ under the regularity
conditions. Then,
\bse
&&N^{-1}\sumI\left\{\frac{r_i}{\pi}\rho(y_i)-\frac{1-r_i}{1-\pi}\right\}\wt\b(\x_i,\bt)\\
&=&N^{-1}\sumI\left\{\frac{r_i}{\pi}\rho(y_i)-\frac{1-r_i}{1-\pi}\right\}\b(\x_i,\bt)\\
&&+\E\left[\left\{\frac{R}{\pi}\rho(Y)-\frac{1-R}{1-\pi}\right\}\E\left\{\wt\b(\X,\bt)-\b(\X,\bt)\mid Y\right\}\right]+o_p(n^{-1/2})\\
&=&N^{-1}\sumI\left\{\frac{r_i}{\pi}\rho(y_i)-\frac{1-r_i}{1-\pi}\right\}\b(\x_i,\bt)+o_p(n^{-1/2}),
\ese
where the second equality holds since $\rho(y)$ is bounded under the regularity condition and $\wt\b(\x,\bt)-\b(\x,\bt)=o_p(1)$, and the last equality holds since $\E\{\wt\b(\X,\bt)-\b(\X,\bt)\mid y\}=\E\{\U(y,\X,\bt)-\U(y,\X,\bt)\mid y\}=\0$.
This fact and \eqref{eq:whbt} lead to
\bse
&&-\left\{\A(\bt)^{-1}+o_p(1)\right\}\sqrt{n}(\wh\bt-\bt)+o_p(1)\\
&=&\sqrt{n}N^{-1}\sumI\frac{r_i}{\pi}\rho(y_i)\U(y_i,\x_i,\bt)-\sqrt{n}N^{-1}\sumI\left\{\frac{r_i}{\pi}\rho(y_i)-\frac{1-r_i}{1-\pi}\right\}\b(\x_i,\bt)+o_p(1)\\
&=&N^{-1/2}\sumI\sqrt{\pi}\left[\frac{r_i}{\pi}\rho(y_i)\{\U(y_i,\x_i,\bt)-\b(\x_i,\bt)\}+\frac{1-r_i}{1-\pi}\b(\x_i,\bt)\right]+o_p(1).
\ese
\hfill$\blacksquare$


\end{supplement}
\end{document}